\documentclass[11pt]{article}
\usepackage{setspace}
\usepackage{amsmath, hyperref, wasysym}
\usepackage{latexsym}
\usepackage{cite}
\usepackage{amssymb}
\usepackage{amsthm}
\usepackage[margin=0.75in]{geometry}
\usepackage{wasysym}
\usepackage{fancyhdr}
\usepackage{graphicx}
\usepackage{times}
\usepackage{color}
\usepackage{multicol}
\usepackage{placeins}
\usepackage{authblk}
\usepackage{arydshln}
\usepackage{float}
\usepackage{dsfont}
\usepackage{caption}
\usepackage{subcaption}
\usepackage{booktabs}
\usepackage{enumerate}
\usepackage{enumitem}
\usepackage{adjustbox}
\restylefloat{table}
\usepackage{lipsum}
\usepackage[labelfont=bf]{caption}

\begin{document}
\title{Modeling stripe formation on growing zebrafish tailfins}
\author[a,*]{Alexandria Volkening}
\author[b]{Madeline R Abbott}
\author[c]{Dorothy Catey}
\author[d]{Neil Chandra}
\author[e]{Bethany Dubois}
\author[f]{Francesca Lim}
\author[g,h]{Bj\"{o}rn Sandstede}
\affil[a]{NSF--Simons Center for Quantitative Biology, Northwestern University, Evanston, IL}
\affil[b]{Biostatistics, University of Michigan, Ann Arbor, MI}
\affil[c]{Emsi, Moscow, ID}
\affil[d]{Facebook, Menlo Park, CA}
\affil[e]{D.E.\ Shaw Research, New York, NY}
\affil[f]{Citizens Bank, Providence, RI}
\affil[g]{Division of Applied Mathematics, Brown University, Providence, RI}
\affil[h]{Data Science Initiative, Brown University, Providence, RI}
\affil[*]{Email address for correspondence: alexandria.volkening@northwestern.edu}
\maketitle

\begin{abstract}
As zebrafish develop, black and gold stripes form across their skin due to the interactions of brightly-colored pigment cells. These characteristic patterns emerge on the growing fish body, as well as on the anal and caudal fins. While wild-type stripes form parallel to a horizontal marker on the body, patterns on the tailfin gradually extend distally outward. Interestingly, several mutations lead to altered body patterns without affecting fin stripes. Through an exploratory modeling approach, our goal is to help better understand these differences between body and fin patterns. By adapting a prior agent-based model of cell interactions on the fish body, we present an \emph{in silico} study of stripe development on tailfins. Our main result is a demonstration that two cell types can produce stripes on the caudal fin. We highlight several ways that bone rays, growth, and the body--fin interface may be involved in patterning, and we raise questions for future work related to pattern robustness.
\end{abstract}


\section{Introduction}
\label{intro}
\begin{figure*}[t]
\centering
\includegraphics[width=0.8\textwidth]{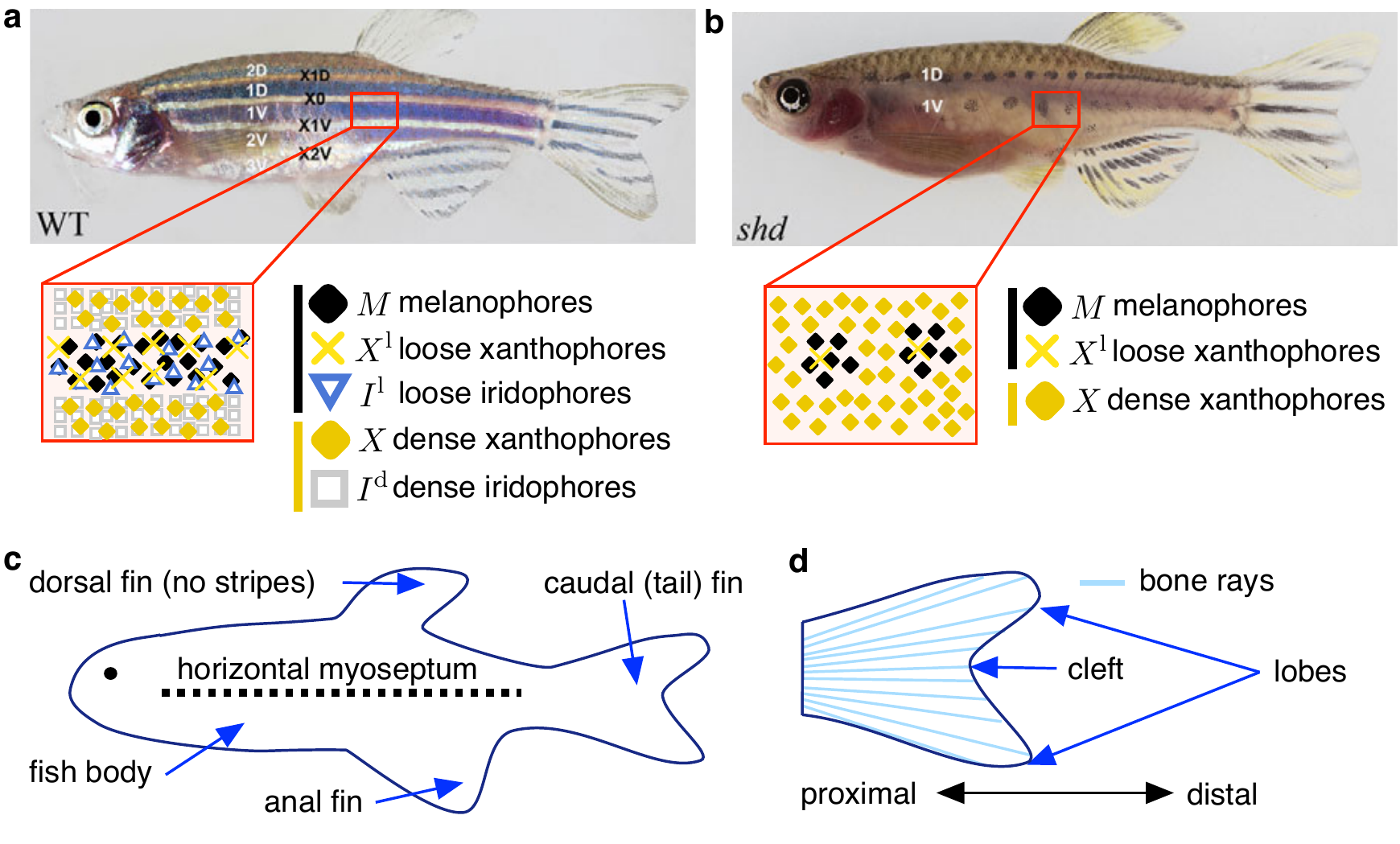}
\caption{Motivation and notation. (a) Wild-type zebrafish feature stripes across their bodies, as well as on their anal and caudal fins. Dark stripes consist of melanophores, loose xanthophores, and loose iridophores; and light stripes are made up of dense xanthophores and dense iridophores \cite{Frohnhofer,hirata2005pigment}. For the remainder of this paper, we follow the empirical convention of calling light stripes ``interstripes'' and dark stripes ``stripes.'' (b) The \emph{shady} mutant (encoding ltk) lacks iridophores \cite{Frohnhofer,Lopes}. While spots form on the bodies of these mutant fish, their fin patterns remain largely unchanged \cite{Frohnhofer}. Although we show \emph{shady} with dense and loose xanthophores as in the model \cite{volkening2}, this is a simplification as \emph{shady} xanthophores seem to be in an intermediate form \cite{2016heterotypic}. (c--d) Here we overview the fish notation relevant to our study. Fish images in (a) and (b) are reproduced with adaption from \cite{Frohnhofer}, licensed under CC-BY $3.0$ (https://creativecommons.org/licenses/by/3.0/), and published by The Company of Biologists, Ltd; we added the red boxes and cell schematics. \label{figfins}}
\end{figure*}

A model organism with extensive biomedical applications, the zebrafish (\emph{Danio rerio}) features black and gold stripes across its body and fins (see Fig.\ \ref{figfins}a). These namesake patterns emerge during development due to the interactions of pigment cells, which self-organize on the growing fish skin. Although it may seem natural for the same cell interactions to drive patterning across the fish, mutants \cite{fadeev2015tight} in which body patterns are altered but fin stripes remain unchanged complicate this picture. The timeline of pattern development is also different on the body and the fins: body stripes appear parallel to an existing horizontal marker \cite{Frohnhofer}, but caudal fin patterns form outward during growth, raising questions about how horizontal alignment is specified there. To better understand these differences, here we develop an agent-based model of pigment cell interactions on growing tailfins.

Three main types of pigment cells make up zebrafish patterns: dark melanophores, orange/yellow xanthophores, and silver/blue iridophores. Until recently, the focus of the biological and mathematical communities was on the first two cell types, and a series of interactions taking the overarching form of short-range activation and long-range inhibition was identified \cite{Nakamasu}. This is a condition of Turing-type pattern formation \cite{Turing,mein}; when it is invoked in zebrafish models, patterns are traditionally viewed as forming due to the reaction and diffusion of melanophores and xanthophores. In the last few years, however, new dynamics have surfaced, and it has become clear that iridophores play a critical role on the zebrafish body \cite{Singh,PatPLos,Jan,Frohnhofer}. This is seen most strikingly in \emph{shady}, a mutant that lacks iridophores and displays black spots across its body in place of stripes \cite{Frohnhofer,Lopes} (see Fig.\ \ref{figfins}b). Empirical studies \cite{Singh,PatPLos,Jan} have shown that iridophores lead patterning and instruct the other cells on the fish body. Affirming this \emph{in silico}, the two-cell model \cite{volkening} could not produce complete body stripes without accounting for iridophores indirectly. These discoveries have led to active debate \cite{Mahalwar,mahalwar2,kondo2015comment}, as researchers work to reconcile the previous picture of two cell populations with new data about iridophores.

\begin{figure}[t]\centering
\includegraphics[width=0.75\textwidth]{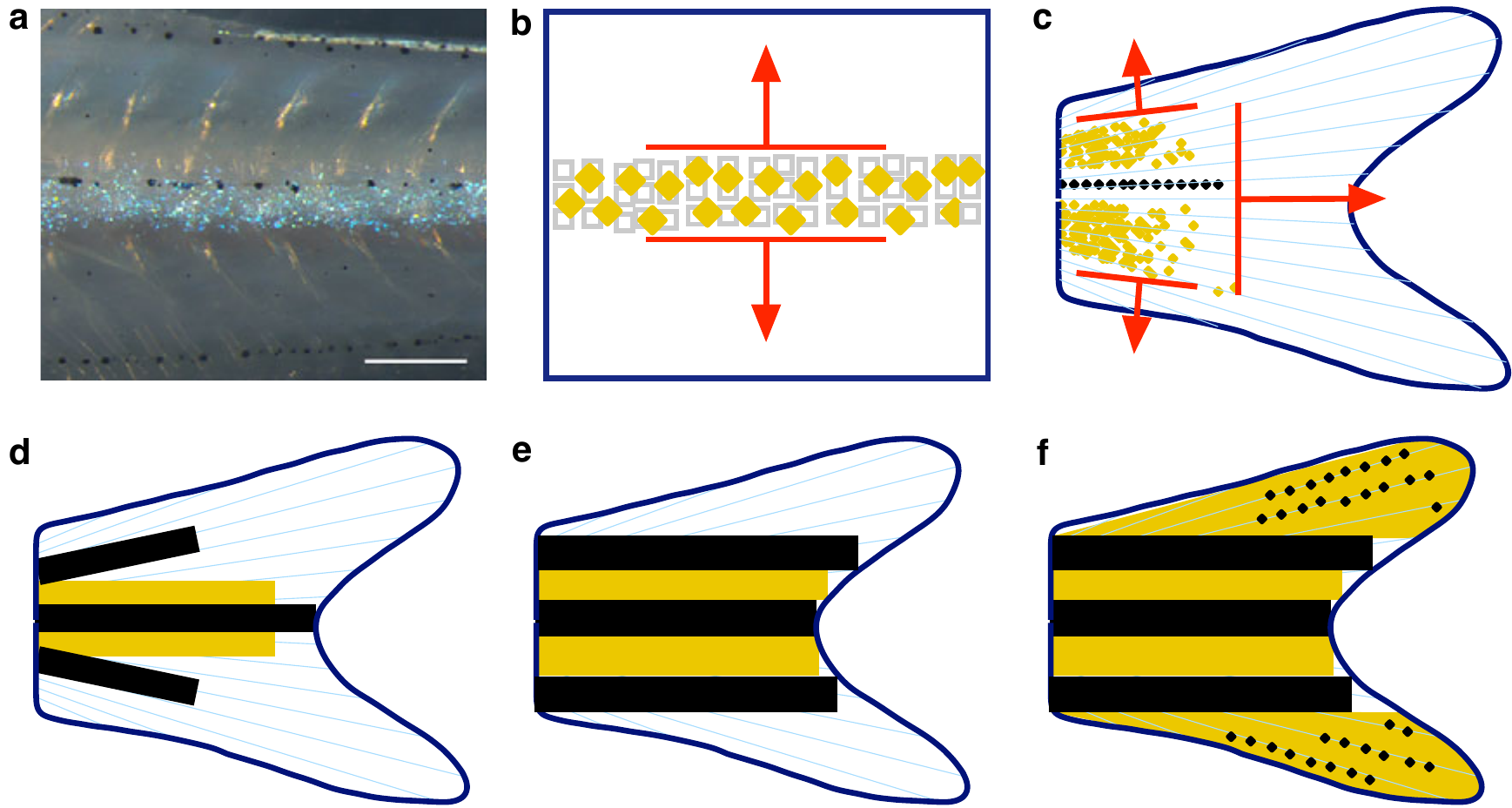}
\caption{\label{ablCartoon} Differences in pattern formation on the body and tailfin. (a) On the zebrafish body, iridophores emerge along the horizontal myoseptum to create the first interstripe \cite{Singh,Jan}. (b) The strip of iridophores at the horizontal myoseptum provides an initial source of alignment, and stripes form sequentially in parallel to this marker \cite{Frohnhofer}. (c) In contrast, new stripes on the tailfin form by slowly extending from the proximal to the distal edge, so that the direction of pattern growth is roughly perpendicular to the developing stripes \cite{tu2010clonal}. (e--f) Based on our observations of empirical images in \cite{Parichy}, stripes initially develop radially (e.g., in association with the bone rays), before adopting a more horizontal alignment on the tailfin. (f) The presence of melanophores aligned with the bone rays is particularly pronounced in the fin lobes (our observation based on images in \cite{Parichy}). Image (a) is reproduced from \cite{Frohnhofer}, licensed under CC-BY $3.0$ (https://creativecommons.org/licenses/by/3.0/), and published by The Company of Biologists, Ltd.}
\end{figure}

As we show in Fig.\ \ref{figfins}b, although \emph{shady} fish develop spots on their bodies, the stripe patterns on their fins remain unchanged \cite{Frohnhofer}. This raises several questions, and one immediately wonders why melanophores and xanthophores alone are incapable of producing stripes on the body, yet are sufficient on the fins. One possible explanation is that their interactions are intrinsically different in these two regions. Opinions on this diverge \cite{nusslein2017fish,mahalwar2,Jan}, with some researchers hypothesizing that melanophores and xanthophores behave largely the same in both domains \cite{kondoTuringQuestion, kondo2015comment}. It is also possible that different cell precursors or micro-environments are at work on the body and tailfin \cite{Spiewak}.

In addition to differences in the types of pigment cells involved in pattern formation on the body and tailfin, the timelines of pattern development in these regions are different. As we show in Fig.~\ref{ablCartoon}a, body stripes form sequentially outward from a central strip of iridophores at the horizontal myoseptum. Loss of the horizontal myoseptum in the \emph{choker} mutant leads to labyrinth patterns \cite{Frohnhofer}. Similarly, disrupting the initial strip of cells in Fig.~\ref{ablCartoon}a through ablation produces patterns with lost directionality \cite{PatPLos,Yamaguchi}. The horizontal myoseptum therefore specifies initial alignment on the body and, by spreading ventrally and dorsally from this marker, iridophores provide further directionality cues \cite{Singh}. Body stripes then emerge sequentially outward from the fish center in a direction parallel to established stripes (see Fig.~\ref{ablCartoon}b). In contrast, as we note in Fig.~\ref{ablCartoon}c, tailfin stripes grow from proximal to distal (e.g., see images of fin development in \cite{Parichy}); proximal-to-distal patterning has also been noted after tailfin amputation \cite{rawls2000zebrafish,eom2015long}. Although new stripes are added dorsally and ventrally, they seem to develop by gradually extending in a direction roughly perpendicular to the stripes on the tailfin. Moreover, as we illustrate in Fig.~\ref{ablCartoon}d--e, stripes initially appear radially along the bone rays, but later adopt a horizontal alignment (our observation based on images in \cite{Parichy}). The presence of melanophores along the bones (illustrated schematically in Fig.~\ref{ablCartoon}f) is particularly pronounced in the lobes (e.g., Fig.~13 in \cite{Parichy}).

While much work has been done to identify cell dynamics on the zebrafish body, there are fewer results focused specifically on uncovering pigment cell behaviors on the fins. On the empirical side, melanophores and xanthophores extracted from fins have been used in \emph{in vitro} experiments \cite{Yamaguchi,Inaba,Inou,kondoFrogEgg}. Tu and Johnson \cite{tu2010clonal} have studied melanophore lineages, precursors, and spreading on fins. Additionally, Parichy \emph{et al.} \cite{Parichy} provide images of stripe formation on zebrafish as a whole. The tailfin is amenable to amputation, and the studies \cite{goodrich,kizil2009simplet,tu2011fate,rawls2000zebrafish,pfefferli2015art,goodrich1954analysis} involved amputating the fin and observing regrowth. Furthermore, several works (e.g., \cite{mills2007deconstructing,quigley2005,Iwashita,mellgren2006pyewacket,ParTur130,eom2012melanophore}) discuss the impact of mutations on the fins of zebrafish or other \emph{Danio}s. Prior mathematical studies \cite{bullara, Gaffney,Cops, Bloomfield, painter, Shinbrot,MorDeutsch,Nakamasu,volkening,volkening2}, however, have either focused on the zebrafish body or made no attempt at differentiating between the body and the fins.

Studying the tailfin is an important problem from the perspective of zebrafish-pattern research because it would provide a more cohesive understanding of pigment cell interactions \cite{kondo2015black,Jan}. The ability of tailfin patterns to regenerate after amputation also suggests potential applications to wound healing and tissue regeneration \cite{Spiewak}. We therefore put forward an initial modeling step toward a more comprehensive view of zebrafish stripes here. Our objective is twofold:
\begin{itemize}[noitemsep,nolistsep]
\item[(1)] to help elucidate whether there are necessarily differences in the cell interactions underlying pattern formation on the body and tailfin of zebrafish; and
\item[(2)] to help determine if patterns can arise autonomously due to melanophore and xanthophore interactions or if external cues are needed to align stripes on the tailfin.
\end{itemize}
Based on experimental observations that iridophores are not involved in stripe formation on the caudal fin \cite{kondoTuringQuestion}, we approach this problem by adapting the prior melanophore-xanthophore body model \cite{volkening} on growing tailfin domains. This model has the overarching form of short-range activation and long-range inhibition \cite{Turing, mein}, and it allows us to test if the same melanophore and xanthophore interactions can account for stripe formation on the body and tailfin. We test the body model \cite{volkening} on caudal-fin domains under different mechanisms of epithelial growth and extend it to account for fin-specific dynamics. Our model suggests that melanophores and xanthophores can produce stripe patterns on the tailfin without iridophores, but additional mechanisms may be needed to ensure stripes form robustly. We find that epithelial growth and bone rays may provide directionality cues, serving as a means of reconciling differences between body and fin patterns on zebrafish.

Regarding manuscript organization, we begin in Sect.~\ref{proposed} by motivating and identifying a series of proposed mechanisms that may be involved in pattern formation on the caudal fin. We then present our models of tailfin growth and cell interactions in Sect.~\ref{modelsec}. By simulating our agent-based model on growing fin domains in Sect.~\ref{results}, we test how skin growth, the body--fin interface, and the presence of bone rays may affect stripe formation and alignment on the tailfin. We conclude in Sect.~\ref{discussion} by summarizing our study and highlighting directions for future work. We include all of our model parameters and detailed simulation conditions in Appendices \ref{appendix:parameters} and \ref{appendix:simulation}.

\section{Proposed mechanisms of pattern formation on the tailfin} \label{proposed}

To gain some intuition into what instructs stripe patterns to form on the tailfin without iridophores, it is useful to compare the timelines of development on the body and caudal fin. As we illustrate in Fig.\ \ref{ablCartoon}a--b, stripes and interstripes develop sequentially in directions dorsal and ventral to a horizontal marker on the fish body \cite{Frohnhofer}. In particular, narrow stripes emerge parallel to the horizontal myoseptum across the full length of the body, and then these stripes widen in time. In contrast, stripes and interstripes gradually develop by spreading from proximal to distal across the caudal fin. (The boundary of un-patterned space is parallel to stripes on the body, but it is perpendicular on the tailfin.) 

Disparate types of epithelial growth on the body and tailfin could potentially explain the differences in pattern development between these two regions. While the skin on the fish body grows uniformly, tailfin growth occurs distally through the addition of new bone segments at ray tips \cite{kondoTuringQuestion,goldsmith2003saltatory,iovine2000genetic,iovine2007conserved}. It has been hypothesized that such ray growth could be a source of directionality on the tailfin \cite{kondoTuringQuestion}. Unlike on the body, however, it is less clear what kind of epithelial growth occurs on the fins \cite{goldsmith2003saltatory}. If skin and bone growth are both limited to the distal edge, skin cells would differentiate in the fin lobes in response to the addition of new bone-ray segments. In this setting, pigment cell positions would not be altered by skin growth, and stripe alignment could be specified by the controlled appearance of empty space at the distal edge (see Fig.~\ref{model}a). Alternatively, under uniform epithelial growth like that present on the body, skin cells would constantly differentiate across the fin (see Fig.~\ref{model}b). Cell positions would evolve as if they were sitting on an expanding rubber sheet, providing an alternative source of radial directionality. Notably, in the work \cite{tu2010clonal}, Tu and Johnson tracked melanophore lineages across the caudal fin. They found that most new melanophores are added in the distal third of the developing stripes (though they also observed some melanophore birth across the entire pattern). These findings \cite{tu2010clonal} could be accounted for by a combination of uniform and distal epithelial growth, for example.

Although the simplest explanation for patterning on the caudal fin may be that skin growth steps in to specify stripe alignment and fill the role of iridophores on the body, empirical images of fins in \cite{Parichy} point to a few features of cell organization that complicate this hypothesis. (We note that we cannot republish images from \cite{Parichy}, but we describe them here and refer the reader to this reference for fin images.) First, the radial nature of melanophore organization on the tailfin is noticeable by eye: melanophores appear dispersed along the bones, particularly in the lobes (see Fig.~\ref{ablCartoon}d and \ref{ablCartoon}f). As the fish develops, stripes gradually spread across the fin, initially radially, but later adopting a parallel arrangement (see Fig.~\ref{ablCartoon}d--f). The intermingling of melanophores and xanthophores \cite{eom2015long} in the lobes stands out, and it is unclear how to reconcile these dynamics with the local competition observed between these cells on the body \cite{Nakamasu,TakahashiMelDisperse}. Lastly, we do not know if any cell interactions occur across the proximal boundary, where the body and fin connect. It is possible that cells enter the tailfin from the body, providing a proximal source of cells and alignment cues from the body pattern (e.g., the early study \cite{goodrich1954analysis} suggested that xanthophore precursors could conceivably enter the caudal fin from the body).

There are multiple mechanisms that could explain the apparent radial organization of melanophores on the tailfin. Bone rays could physically corral cells and affect their movement, especially where the fin is thinnest near the lobes. Alternatively, melanophores may differentiate from precursors associated with or between the rays. For example, \cite{tu2010clonal} showed that the boundaries of melanophore-clone distributions in the tailfin are parallel to the bone rays. In an early study \cite{goodrich1954analysis}, melanophores were observed as entering interstripes along fin rays on the anal fin. In this case, bones could serve as an external source of directionality on the tailfin like the horizontal myoseptum does on the body. The initially radial, later horizontal, development of stripes could be related to fin thickness or be explained by an additional signal spreading from proximal to distal on the tailfin. Finally, it is conceivable that the melanophores that we observe along the rays in images from \cite{Parichy} are in a larval (e.g., inactive) form and are later replaced by adult cells. The presence of distinct classes of melanophores on the fins has been noted by several studies (e.g., \cite{tu2010clonal,rawls2000zebrafish}).

To put some delimiters on the realm of possible processes at work on the tailfin, we test the following mechanisms for wild-type pattern development:
\begin{itemize}[noitemsep,nolistsep]
\item[{I}] distal epithelial growth (Fig.~\ref{model}a);
\item[{II}] alignment cues from body patterns (Fig.~\ref{model}g);
\item[{III}] uniform epithelial growth along bone rays (Fig.~\ref{model}b);
\item[IV] melanophore migration along bone rays (Fig.~\ref{model}e); and
\item[{V}] melanophore differentiation in association with bone rays (Fig.~\ref{model}h).
\end{itemize}
In all cases, we consider the interactions of two cell types (melanophores and xanthophores) on growing tailfin domains. To implement Mechanisms I--V, we introduce a model of cell interactions that includes switch parameters (see Sect.~\ref{modelsec}). By setting our switch parameters to extreme values (e.g.,$-1$ or $1000$ cells), we select the mechanisms that we include in a given simulation.

\section{Model}\label{modelsec}

On the caudal fin, stripes are made up of a layer of xanthophores sandwiched between two sheets of melanophores, and interstripes contain a sheet of xanthophores \cite{hirata2005pigment}. Although fins also contain iridophores, they do not match up with the stripes like on the body \cite{kondoTuringQuestion}. Recent work \cite{Mahalwar,walderich2016homotypic} has shown that body xanthophores appear in fundamentally different forms in stripes and interstripes, and we expect that a similar distinction is present on the tailfin. In this first model of fin patterns, we therefore take a minimal approach and consider a single layer of cells, with black melanophores ($M$) in stripes and gold dense xanthophores ($X$) in interstripes.

We base our model for cell interactions on a prior model \cite{volkening} of $M$ and $X$ cells on the zebrafish body. We use an agent-based approach, modeling cells as point masses that undergo deterministic migration (by coupled ordinary differential equations) and stochastic birth and death (through noisy, discrete-time rules). In particular, we let
\begin{align*} \textbf{M}_i(t) &= \text{position of the center of the $i$th melanophore at time $t$,} \\
\textbf{X}_j(t) &= \text{position of the center of the $j$th xanthophore at time $t$,} \\
N_\text{M}(t) &= \text{number of melanophores at time $t$, and } \\
N_\text{X}(t) &= \text{number of xanthophores at time $t$},
\end{align*}   
where we measure time in days. The number of cells on the fin grows in time, varying from roughly $250$ to $5,000$ cells over the course of one of our simulations.

We track the full tailfin domain, with its growing shape determined by tracing curves around images in \cite{Parichy} and interpolating between these curves. Using images in \cite{Parichy}, we also specify bone rays on our domains. To test how the presence of bones may affect pattern formation (namely, Mechanisms III--V in Sect.~\ref{proposed}), we discretize the bone rays as below:
\begin{align}
\{\textbf{B}^j_i(t)\}_{i=1,\ldots,500} &= \text{ set of coordinates for the $j$th discretized bone ray at time $t$,} \label{bone}
\end{align}
where the initial coordinate of each discretized bone is along the $y$-axis and the discretization step in the $x$-direction is $\Delta b_x = x_\text{max}(t)/499$ (as we show in Fig.~\ref{finGrowth}, $x_\text{max}(t)$ is the length of the caudal fin at time $t$ days). (When we plot bones in our simulations, we only show the discretized bone coordinates that fall within our tailfin domains.) We use the same number of discretization points per bone regardless of fin size; in future models, one could alternatively keep $\Delta b_x$ constant in time.

We simulate pattern development from when zebrafish are roughly $18$ days post fertilization (dpf) to when they reach adult stages at $150$ dpf. It is important to note that growth rates vary \emph{in vivo}, and experimentalists prefer to track time using developmental stage or a measure of body length called standardized standard length (SSL) \cite{Parichy}. Generally reported without units \cite{McMen2016}, SSL is a measurement of standard length (SL) in mm based on representative zebrafish; see Fig.~\ref{finGrowth}a. To follow conventions and account for broad audiences, we track time using both developmental stage and time in dpf (see Sect. \ref{growth} for details). We provide our relationships between SSL, developmental stage, and time in Table \ref{table:SSL}.

     \begin{table}[t]
     \centering
    \begin{tabular}{l l  l l}
    {\bf{Stage}} & {\bf{SSL}} & {\bf{Estimated age}}  & {\bf{Fin image in \cite{Parichy}}}\\ 
    \toprule
     Pelvic fin bud (PB) & $7.2$ & $18$ dpf & Fig.\ $47$G'\\ \midrule
     Pelvic fin ray (PR) & $8.6$  & $24$ dpf & Fig.\ $49$G' \\ \midrule
     Squamation onset posterior (SP) & $9.8$& $33$ dpf &Fig.\ $51$G' \\ \midrule
     Juvenile (J) & $11.0$ & $43$ dpf & Fig.\ $53$G'\\ \midrule
     Juvenile$+$ (J$+$) & $13.0$ & $70$ dpf & Fig.\ $54$G'\\ \midrule
     Juvenile$++$ (J$++$) & $16.0$ & $93$ dpf & Fig.\ $55$E'\\   \midrule
     Adult (A) & $26$ &$278$ dpf & Fig.\ $56$E'\\   \bottomrule
     \end{tabular}
     \caption{{Stages of development, estimated fish age, and fin images. As we show in Fig.~\ref{finGrowth}a, SL is a measurement of fish body length (not including the caudal fin). SSL is a characteristic SL based on representative zebrafish \cite{Parichy}. SSL is conventionally not reported with units \cite{McMen2016}, but it is associated with SL measured in mm. To produce our growing domains, we trace fin boundaries in fish images from \cite{Parichy}. We estimate age from SSL using Eqn. (\ref{dfp})}} \label{table:SSL}
     \end{table}

In the following subsections, we describe our model of growing fin domains (Sect.~\ref{growth}) and then introduce our agent-based model of cell interactions, adapted from the body model \cite{volkening} (Sect.~\ref{sec:Brief Model}). For a summary of our model, see Fig.~\ref{model}. We include further details about our simulations and parameters for reproducibility in Appendices \ref{appendix:parameters} and \ref{appendix:simulation}.

\subsection{Model of growing tailfin domains}\label{growth}

\begin{figure*}[tbhp]\centering
\includegraphics[width=0.75\textwidth]{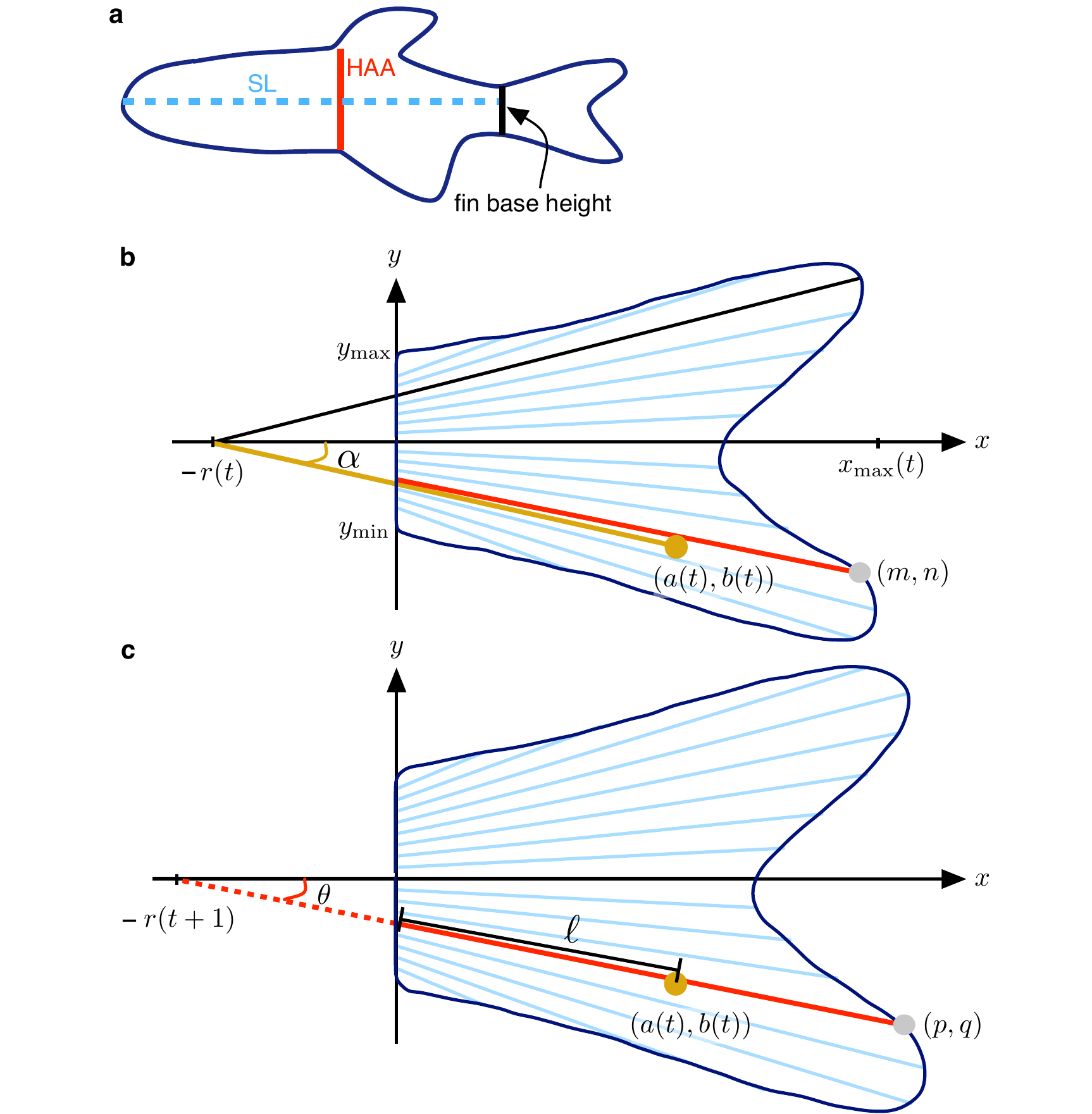}
\caption{Fin domain and growth. (a) We develop our growing domain by tracing images of fins and bone rays in \cite{Parichy}; also see Table \ref{table:SSL}. To determine the right spatial scaling for these images, we use the SSL reported in \cite{Parichy} for each image. By relating this SSL to measurements of body length (SL) and height (HAA), we estimate the height of the fin at its proximal edge (see Eqn. (\ref{HAA})). The height that we estimate for the fin base provides the necessary scaling information for our domains. (b--c) When we include uniform epithelial growth (Mechanism III in Sect.~\ref{proposed}), we scale cell positions in time with bone growth; also see Fig.~\ref{model}b. For example, to adjust the position of the cell at position $\textbf{X}_i(t)= (a(t),b(t))$, we first identify the scaling angle ($\alpha$) and the rays (highlighted in red) that are closest to $\textbf{X}_i(t)$ in the current fin domain at day $t$ and the next fin domain at day $t+1$. To determine the amount to scale in the direction $\alpha$, we calculate how much the nearest rays grow between these two days. Here $(m,n)$ and $(p,q)$ are the endpoints of the closest rays to $(a(t),b(t))$ at day $t$ and $t+1$, respectively. Moreover, $\ell \approx \left| \frac{a(t)}{\cos (\theta)} \right|$; we note that this is generally not a strict equality because we define $-r(t+1)$ to be the position where the third most dorsal bone at day $t+1$ intersects the $x$-axis, and the bones do not all intersect the $x$-axis at exactly the same point (see Sect.~\ref{growth} for details). \label{finGrowth}}
\end{figure*}

Because it contains many caudal fin images, we draw heavily on the work \cite{Parichy} to specify the shape, bone-ray placement, and growth rates for our domains. In particular, we manually sharpen seven caudal-fin images from {\cite{Parichy}} using Adobe Photoshop, identify the $(x,y)$ coordinates of the fin boundaries in the cleaned images using Matlab, and smooth any rough edges in the resulting curves. As we note in Table \ref{table:SSL}, these fin images span from $7.2$ SSL to $26$ SSL. To approximate the fish age and specify scale bars for each image, we convert these SSL (body length) measurements to HAA (body height) measurements as follows:
\begin{align}
\text{HAA} &= 0.259  \times \text{SSL} - 0.985 \text{ mm}, \label{HAA}
\end{align}
where this relationship is our approximation from figures in \cite{Parichy}. Based on images in \cite{Parichy}, we then estimate the height of the tailfin at its proximal boundary as roughly equal to $70\%$ of the fish body height (that is, $\text{our fin base height} = 0.7 \times \text{HAA}$). See Fig.\ \ref{finGrowth}a. To approximate the age of the fish in each image, in turn, we relate SSL and time as below:
\begin{align}
\text{time in dpf } &=
\begin{cases}
   \exp \left(\frac{4.8291 + \text{SSL}}{4.1994}\right) & \text{  if time $\le 85$ dpf} \\
  \frac{10.949 + \text{SSL}}{0.0541} & \text{  if time $> 85$ dpf} \label{dfp} \\
\end{cases},
\end{align}
where this relationship is our approximation based on graphs in \cite{Parichy}. Through this process, we assign a spatial scale and time stamp (in dpf) to each of our traced caudal-fin images (see Table \ref{table:SSL}). To span the time between our seven images, we calculate a boundary curve for each day of simulated development by assuming linear growth and a continuous transformation to each new image.

In addition to defining the outline of our fin domains, we add bone rays each day. Based on the images that we gathered from \cite{Parichy}, we estimate that the angle between consecutive bone rays is roughly $2.5$--$4.5^{\circ}$, decreasing with zebrafish age (Fig.\ \ref{finGrowth}b). We specify bone rays at a constant $2.833^{\circ}$ apart, and, motivated by \cite{rolland2012morphogen, iovine2007conserved,goldsmith2006developmental}, we include $18$ rays on our domain. Each day we define the bones in five steps:
\begin{enumerate}[noitemsep,nolistsep]
\item Find the coordinates of the upper and lower endpoints of the proximal boundary of the fin, namely $(0,y_{\text{max}}(t))$ and $(0,y_{\text{min}}(t))$.
\item Define $y_{\text{avg}} = \frac{y_{\text{max}(t)} + y_{\text{min}(t)}}{2}$.
\item Calculate the limits of ray placement by shrinking the interval slightly so that bones do not emanate directly from the upper and lower boundary of the domain.
\begin{align*} y^\text{bone}_\text{max} &= (y_\text{max}(t) - y_\text{avg})\times 0.95 + y_\text{avg},\\
y^\text{bone}_\text{min} &= (y_\text{min}(t) - y_\text{avg}) \times 0.95 + y_\text{avg}.
\end{align*} 
\item Discretize the proximal fin boundary into $18$ equidistant points between $(0,y^\text{bone}_{\text{max}}(t))$ and $(0,y^\text{bone}_{\text{min}}(t))$.
\item Define rays emanating from each of these points at angles ranging from $25.5^{\circ}$ to $-22.667^{\circ}$ with a step of $2.833^{\circ}$.
\end{enumerate}

In some simulations, we include uniform epithelial growth (Mechanism III in Sect.~\ref{proposed}), and so we scale cell positions with bone growth (see Fig.~\ref{model}a--b). For a given cell, we implement this by first identifying the direction in which we will scale the cell coordinates and then determining the scaling factor that we will apply (see Fig.~\ref{finGrowth}b--c). We make the simplifying assumption that each day all the rays intersect at the same point $(-r(t),0)$. We define this point $(-r(t),0)$ to be the place where the third most dorsal ray (highlighted in black in Fig.~\ref{finGrowth}b) crosses the $x$-axis. In particular, for a cell with coordinates $(a(t),b(t))$ at time $t$, its scaling angle $\alpha$ is 
\begin{align*} \alpha &= \text{ scaling angle } = \text{ atan}\left( \frac{b(t)}{a(t) + r(t)} \right).
\end{align*}
We then calculate the cell's scaling factor by finding its nearest bone rays at time $t$ and $t+1$ days (e.g., they have endpoints $(m,n)$ and $(p,q)$ in the example in Fig.~\ref{finGrowth}b--c):
\begin{align*} S = \text{scaling factor} = \frac{\text{length of closest ray to $(a(t),b(t))$ at time $t+1$}}{\text{length of closest ray to $(a(t),b(t))$ at time $t$}}. 
\end{align*}
We estimate ray length using the discretized bone points in Eqn.~\ref{bone} that fall within our domain. Lastly, we update the coordinates of the cell at $(a(t),b(t))$ as below:
\begin{align*} a(t) &= S \left| \frac{a(t)}{\cos (\theta)}\right| \cos (\alpha), \\
b(t) &= (S-1)  \left| \frac{a(t)}{\cos (\theta)} \right| \sin (\alpha) + b(t),
\end{align*}
where $\theta = \text{ atan} \left(\frac{q}{p+r(t)} \right)$ is the angle that the closest ray to the pigment cell on the updated domain makes with the $x$-axis. We note that $\left| \frac{a(t)}{\cos (\theta)} \right| \approx \ell$ in Fig.~\ref{finGrowth}c.

\subsection{Model of cell interactions}\label{sec:Brief Model}

\begin{figure*}[t!]\centering
\includegraphics[width=0.9\textwidth]{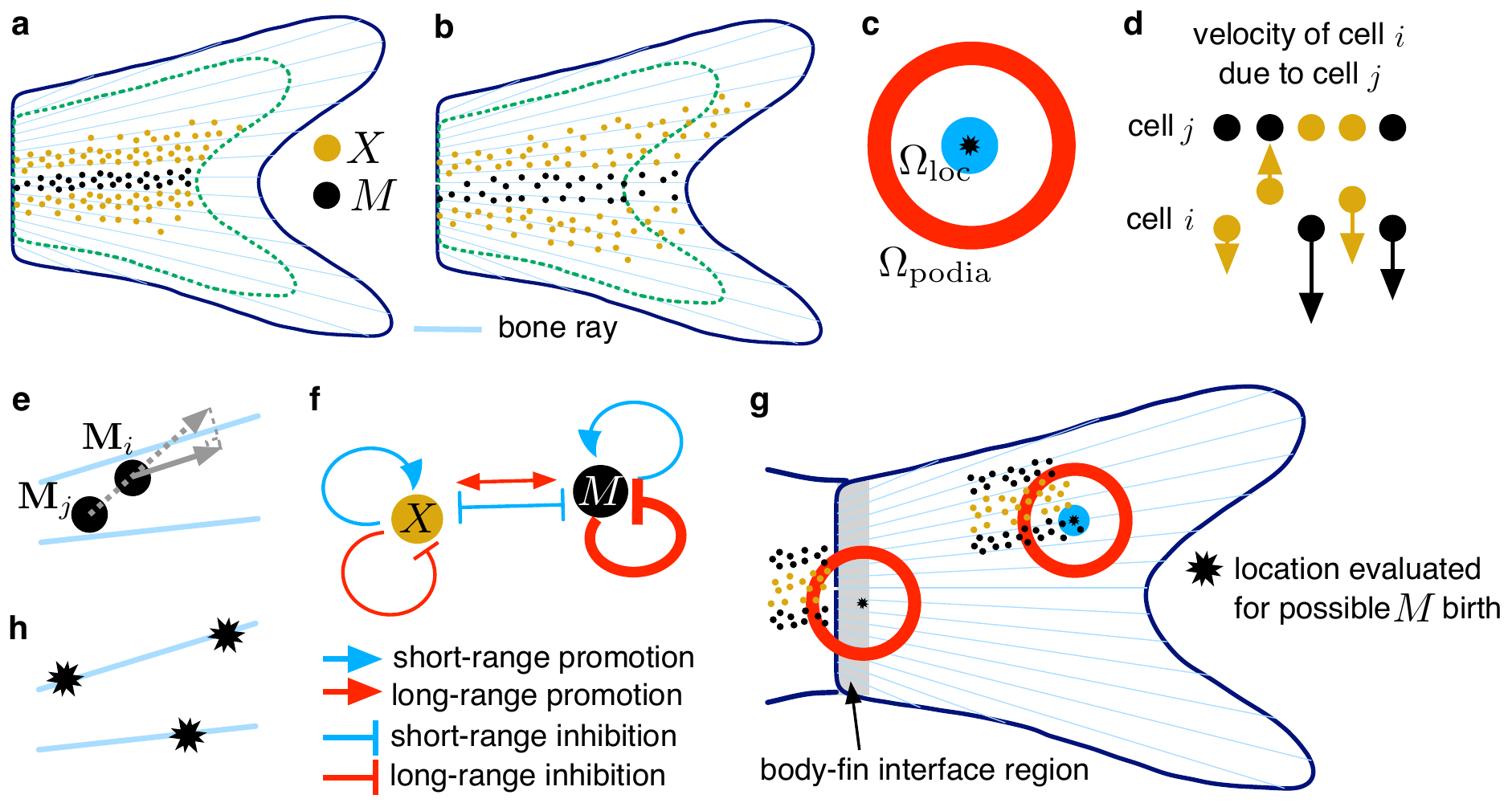}
\caption{Model overview. (a) Under the assumption of distal epithelial growth (Mechanism I in Sect.~\ref{proposed}), we do not scale cell positions as the fin grows. The dashed green curve indicates the fin boundary at time $t$ days and the solid blue curve denotes the boundary at $t+1$ days. (b) In contrast, under uniform epithelial growth (Mechanism III), we scale cell positions along the rays with fin growth (see Fig.\ \ref{finGrowth}b--c and Sect.~\ref{growth} for details). (c) Cell interactions depend on a combination of short- and long-range signals \cite{Nakamasu}. Using the same approach as \cite{volkening}, we account for these interactions through stochastic rules that depend on the number of cells in a short-range disk ($\Omega_\text{loc}$) and a long-range annulus ($\Omega_\text{podia}$) around the cell (or precursor) of interest. (d) Similar to \cite{volkening}, we specify repulsive and attractive forces between cells. The cell separation and arrow lengths in this schematic are not to scale, but capture qualitative features of cell migration (e.g., we indicate that $X$--$X$ repulsion is weaker than $M$--$M$ repulsion with a shorter velocity arrow). As we discuss in Sect.~\ref{movement}, in some simulations we remove the local attractive force from $M$ cells on $X$ cells. (e) Under Mechanism IV, we specify that $M$ cells do not migrate across rays. To implement this for the cell at position $\textbf{M}_i$, we find its velocity and then project this along its nearest bone. (f) Cell birth and death take the form of short-range activation and long-range inhibition in our model and \cite{volkening}. (g) Under Mechanism II, we test how cells receiving alignment cues from developing body stripes may impact patterning. To do this, we adjust our rules for cell birth in a small (grey) neighborhood of the fin's proximal edge. In this neighborhood, we specify that birth depends only on long-range signals. (h) To test how specifying $M$ birth in association with the bones impacts pattering, we limit $M$ birth to locations on the rays under Mechanism V.
\label{model} }
\end{figure*}

Pigment cells on the skin of zebrafish interact locally with neighboring cells (e.g., through direct contact \cite{walderich2016homotypic} or short dendrites \cite{Inaba}) and at long range with cells in neighboring stripes \cite{Nakamasu}. Long-range interactions may be mediated by cellular extensions, such as airnemes or pseudopodia, which have been measured to be up to half a stripe width in length \cite{eom2015long,delta}. Diffusion of signalling factors may also play a role \cite{PatNcomm}. We account for these various length scales (see Fig. \ref{model}c) with four interaction neighborhoods:
\begin{align} \Omega^\textbf{z}_{\text{podia}} &= \text{annulus of inner radius $318$ $\mu$m and width $25$ $\mu$m centered at \textbf{z}}, \label{eq:podia} \\
\Omega^\textbf{z}_\text{rand}  &= \text{disk centered at \textbf{z} with radius $\Delta_\text{rand} = 100$ $\mu$m, } \label{eq:rand} \\
\Omega^\textbf{z}_\text{crowd}  &= \text{local overcrowding disk centered at \textbf{z} with radius $\Delta_\text{XM} = 82$ $\mu$m, and} \label{eq:crowd} \\
\Omega^\textbf{z}_{\text{loc}} &= \text{short-range disk of radius $d_\text{loc} \in \{75,\Delta_\text{XM}\}$ $\mu$m centered at $\textbf{z}$,} \label{eq:loc}
\end{align}
where $\Delta_\text{XM}$ is the average distance between $M$ and $X$ cells at stripe--interstripe boundaries \cite{TakahashiMelDisperse}. Our rules for cell interactions depend on the proportions of cells in these neighborhoods. We implement this by introducing indicator functions that identify whether or not a cell is in a neighborhood around the point $\textbf{z}$:
\begin{align} \mathds{1}_{\Omega^\textbf{z}_\text{region}}(\textbf{C}) &= 
\begin{cases}
1 & \text{ if position $\textbf{C}$ lies within the region centered at $\textbf{z}$}  \\
 0 & \text{ otherwise}
 \end{cases}, \label{indicator}
\end{align}
where region $\in \{\text{podia},\text{rand},\text{crowd},\text{loc}\}$. We note that $\Omega_\text{podia}$, $\Omega_\text{rand}$, and $\Omega_\text{crowd}$ are the same regions used in the body model \cite{volkening}. As discussed there, $\Omega_\text{podia}$ accounts for cell interactions across half a stripe width in distance. In \cite{volkening}, $d_\text{loc} = 75$ $\mu$m for $\Omega_\text{loc}$, but we increase this parameter slightly to $\Delta_\text{XM}$ in some of our fin simulations (see Appendix~\ref{appendix:figure}).

In the next subsections, we describe how we adapt the model \cite{volkening} to account for cell movement (Sect.~\ref{movement}), differentiation (Sect.~\ref{birth}), and competition (Sect.~\ref{death}) on the caudal fin. For details about the order in which we implement these interactions, see Appendix \ref{appendix:flow}. Briefly, we implement $1/\Delta t_\text{mig,birth}$ iterations of migration followed by birth; we then evaluate cells for possible death with a time step of $\Delta t_\text{death} = 1$ day; and finally we implement skin growth with a time step of $\Delta t = 1$ day (note that $\Delta t_\text{mig,birth} \le 1$ day). We provide all of our parameters in Tables \ref{table:mig}, \ref{table:birth}, and \ref{table:death} in Appendix~\ref{appendix:parameters}. With small adjustments in some simulations, these parameters are based on the prior body model \cite{volkening}.

\subsubsection{Cell migration} \label{movement}

For each day of simulated development, we implement $1/\Delta t_\text{mig,birth}$ cycles of cell movement, where $\Delta t_\text{mig,birth} \le 1$ day. We specify the movement of the $i$th $X$ cell, located at position $\textbf{X}_i(t)$, with the same first-order ordinary differential equation used in the prior body model \cite{volkening}, namely:
\begin{align} 
 \frac{d\textbf{X}_i}{dt} &= -\sum_{j=1,j\neq i}^{N_\text{X}} \triangledown Q^\text{XX}(\textbf{X}_j - \textbf{X}_i) - \sum_{j=1}^{N_\text{M}} \triangledown Q^\text{MX}(\textbf{M}_j - \textbf{X}_i), \label{eq:x}
\end{align} 
where the potentials $Q^{pq}$, which describe the effect of cell type $p$ on cell type $q$, have the general form:
\begin{align} Q^{pq}(\textbf{d}) &=  R^{pq} e^{-|\textbf{d}|/r_{pq}} - A^{pq} e^{-|\textbf{d}|/a_{pq}}. \label{eq:pot}
\end{align}
We list our parameter values for cell migration in Table~\ref{table:mig} and provide a qualitative illustration of our cell--cell forces in Fig.~\ref{model}d. To keep cells in the domain, we specify wall-like boundary conditions on our fin-outline curves (see Appendix \ref{appendix:boundary}).

Under Mechanism IV in Sect.~\ref{proposed}, we test whether $M$ migration is affected by bone rays on the tailfin. Dependent on a switch parameter $\zeta$, we specify that the velocity of the $M$ cell at position $\textbf{M}_i$ may be projected along its nearest bone ray, as follows:
\begin{align}
\textbf{V}_i &= -\sum_{j=1,j\neq i}^{N_\text{M}} \triangledown Q^\text{MM}(\textbf{M}_j - \textbf{M}_i) - \sum_{j=1}^{N_\text{X}} \triangledown Q^\text{XM}(\textbf{X}_j - \textbf{M}_i), \label{eq:v} \\
  \frac{d\textbf{M}_i}{dt}&= 
\begin{cases}
    ||\textbf{V}_i|| \textbf{b}_i, &\text{if }\sum_{j=1}^{N_\text{X}}{ \mathds{1}}_{\Omega^{\textbf{M}_i}_{\text{crowd}}} (\textbf{X}_j) + \sum_{j=1}^{N_\text{M}}{ \mathds{1}}_{\Omega^{\textbf{M}_i}_{\text{crowd}}} (\textbf{M}_j )< \zeta \\
    \textbf{V}_i,              & \text{otherwise}
\end{cases}, \label{eq:m}
\end{align}
where $\mathds{1}_{\Omega^{\textbf{M}_i}_\text{crowd}}(\cdot)$ is the indicator function in Eqn.~(\ref{indicator}) and $\textbf{b}_i$ is the unit vector in the direction of the bone ray closest to the $i$th $M$ cell. (We identify the ``closest bone ray'' using the discretized bone points in Eqn.~(\ref{bone}). In particular, across the set of $500$ discretized points representing each of our $18$ bone rays, we define the closest bone to the $i$th $M$ cell as the index $k$ for which $||\textbf{M}_i - \textbf{B}^k_j||$ is smallest.) 

The parameter $\zeta$ serves as a switch in Eqn.~(\ref{eq:m}) that allows us to test Mechanism~IV (Sect.~\ref{proposed}) in some simulations:
\begin{itemize}[noitemsep,nolistsep]
\item \emph{Control case (body-model migration):} When $\zeta = -1$ cells (a condition that can never be met), we specify $M$ migration in the same way as in the prior body model \cite{volkening}.
\item \emph{Mechanism IV:} When $\zeta = 1000$ cells (a value larger than the number of cells that could be present in $\Omega_\text{crowd}$), we project $M$ movement along bone rays.
\end{itemize}

As we show in Fig.~\ref{model}d, we use different parameters for the four potentials in Eqns.\ (\ref{eq:x}) and (\ref{eq:v}). With one exception ($Q^\text{MX}$), these potentials are purely repulsive. Following \cite{volkening}, we specify that $M$ cells are repelled from both $X$ and other $M$ at short range. Xanthophores, in turn, are repelled from other $X$ cells at short range. This is motivated by \emph{in vitro} studies \cite{Inaba, Yamanaka2014} showing that $M$ extracted from fins move away from $X$ cells. We also include repulsive interactions between like cells to maintain experimentally-measured cell--cell distances \cite{TakahashiMelDisperse}. Xanthophores extracted from the fins seem to be attracted to $M$ \emph{in vitro} \cite{Yamanaka2014}. When we include this so-called ``chase-run behavior'' in our model, we account for it in the same way as in the body model \cite{volkening}: we specify that $M$ cells attract $X$ cells very locally but repel $X$ at short range.

Although $Q^\text{XX}$, $Q^\text{MM}$, and $Q^\text{XM}$ describe short-range repulsion in a manner qualitatively similar to \cite{volkening}, we adjust the parameters in these potentials in some simulations, as we note in Table \ref{table:mig}. We consider two types of simulation: in the first type, all of our parameters are the same as those in \cite{volkening} (with the caveat that we discuss in the caption of Table~\ref{table:mig}). In the second type, we adjust the parameters slightly from their body equivalents. These adjustments (namely increasing the length scale over which repulsion occurs and reducing its strength) help widen stripes under Mechanism I when skin growth does not stretch cell positions (as it does on the body). Additionally, when testing Mechanisms IV and V, we remove the attractive component of $Q^\text{MX}$ and instead model $X$ cells as simply repelled from $M$. We made this choice to simplify parameter fitting in this first fin model, since it is less clear how changes made to the migration parameters will impact cell movement when interactions are both repulsive and attractive. Importantly, it was shown in \cite{volkening} that removing attraction in $Q^\text{MX}$ does not strongly impact pattern formation on the body.

\subsubsection{Cell differentiation} \label{birth}

For each day of simulated development, we implement $1/\Delta t_\text{mig,birth}$ cycles of cell birth, where $\Delta t_\text{mig,birth} \le 1$ day varies in our simulations and is chosen so that cell birth keeps pace with fin growth. For each such cycle, we implement birth by first selecting a number of possible locations randomly in the domain. We then evaluate these locations for birth based on noisy rules. Because the tailfin grows significantly over the developmental period we simulate, we specify that the number of locations selected grows in time. In particular, the initial values of $n^\text{M}_\text{diff}$ and $n^\text{X}_\text{diff}$ at $18$ dpf vary in our simulations (see Appendix~\ref{appendix:figure}), but they both grow by $20$ cells each day:
\begin{align*}
N^\text{M}_\text{diff}(t) &= n^\text{M}_\text{diff} + 20 (t - 18) = \text{ number of locations evaluated for $M$ birth on day $t$}, \\
N^\text{X}_\text{diff}(t) &= n^\text{X}_\text{diff}+ 20(t - 18) = \text{ number of locations evaluated for $X$ birth on day $t$}.
\end{align*}
We describe how we identify possible positions for cell birth within our fin domain in detail in Appendix~\ref{appendix:cellBirth}. Briefly, we select these positions uniformly at random from a rectangle surrounding the fin, and we then consider only those locations that fall within the fin. Furthermore, under Mechanism V (see Sect.~\ref{bonebirth}), we model $M$ cells as arising from precursors on the bone rays. In this case, we choose locations for $M$ birth in association with the coordinates of our discretized bones in Eqn.~(\ref{bone}).

Our rules for cell birth at randomly selected positions are adapted from the body model \cite{volkening}, and they depend on the proportions of cells in the short- and long-range neighborhoods in Eqns.~(\ref{eq:podia}--\ref{eq:loc}). In \cite{volkening}, a new $M$ cell appears at the selected location $\textbf{z}$ (if not overcrowded) when there are more $M$ than $X$ in $\Omega^\textbf{z}_\text{loc}$ and more $X$ than $M$ in $\Omega^\textbf{z}_\text{podia}$. ($X$ birth occurs under the opposite conditions; also see Fig.~\ref{model}f.) Here we adjust these base rules to include a switch parameter that allows us to test Mechanism II (the presence of alignment cues from developing body stripes) in some simulations.

We specify that a new $X$ cell appears at the selected location $\textbf{z}  = (z_x,z_y)$ when
\begin{align} & \biggl[ \biggl(\underbrace{z_x < d}_\text{[\emph{A}] Mechanism II switch} \hskip0.3cm 
\text{and} \hskip0.3cm  \underbrace{\sum_{i=1}^{N_\text{X}}{ \mathds{1}}_{\Omega^{\textbf{z}}_{\text{rand}}} (\textbf{X}_i) + \sum_{i=1}^{N_\text{M}}{ \mathds{1}}_{\Omega^{\textbf{z}}_{\text{rand}}}(\textbf{M}_i) = 0}_\text{[${B}$] Mechanism II only impacts birth at low density} \biggr) \text{or}  \hskip0.3cm \underbrace{\sum_{i=1}^{N_\text{X}}{ \mathds{1}}_{\Omega^{\textbf{z}}_{\text{loc}}} (\textbf{X}_i) > \phi \sum_{i=1}^{N_\text{M}}{ \mathds{1}}_{\Omega^{\textbf{z}}_{\text{loc}}}(\textbf{M}_i)}_\text{[\emph{C}] short-range activation}\biggl] \nonumber\\
& \hskip0.3cm \text{and} 
\hskip0.4cm \underbrace{\sum_{i=1}^{N_\text{M}}{ \mathds{1}}_{\Omega^{\textbf{z}}_{\text{podia}}} (\textbf{M}_i) > \psi \sum_{i=1}^{N_\text{X}}{ \mathds{1}}_{\Omega^{\textbf{z}}_{\text{podia}}}(\textbf{X}_i)}_{\text{[\emph{D}] long-range inhibition}} \hskip0.3cm \text{and} \hskip0.3cm  \underbrace{\sum_{i=1}^{N_\text{X}}{ \mathds{1}}_{\Omega^{\textbf{z}}_{\text{crowd}}} (\textbf{X}_i) + \sum_{i=1}^{N_\text{M}}{ \mathds{1}}_{\Omega^{\textbf{z}}_{\text{crowd}}}(\textbf{M}_i) < \kappa}_\text{[\emph{E}] condition to prevent overcrowding}, \label{eq:xbirth}
\end{align}
where we define the indicator functions ${ \mathds{1}}_{\Omega^{\textbf{z}}_{\text{rand}}}(\cdot)$, ${ \mathds{1}}_{\Omega^{\textbf{z}}_{\text{loc}}}(\cdot)$, ${ \mathds{1}}_{\Omega^{\textbf{z}}_{\text{podia}}}(\cdot)$, and ${ \mathds{1}}_{\Omega^{\textbf{z}}_{\text{crowd}}}(\cdot)$ in Eqn.~(\ref{indicator}), and we provide our parameters in Table~\ref{table:birth}.

Similarly, we add a new $M$ cell to the domain at position \textbf{z} when
\begin{align} &\biggl[\biggl(\underbrace{z_x < d}_\text{[${A}$] Mechanism II switch} \hskip0.3cm \text{and} \hskip0.3cm  \underbrace{\sum_{i=1}^{N_\text{X}}{ \mathds{1}}_{\Omega^{\textbf{z}}_{\text{rand}}} (\textbf{X}_i) + \sum_{i=1}^{N_\text{M}}{ \mathds{1}}_{\Omega^{\textbf{z}}_{\text{rand}}}(\textbf{M}_i) = 0}_\text{[${B}$] Mechanism II only impacts birth at low density} \biggr) \text{or}  \hskip0.3cm \biggl( \underbrace{\sum_{i=1}^{N_\text{M}}{ \mathds{1}}_{\Omega^{\textbf{z}}_{\text{loc}}} (\textbf{M}_i) > \alpha \sum_{i=1}^{N_\text{X}}{ \mathds{1}}_{\Omega^{\textbf{z}}_{\text{loc}}}(\textbf{X}_i)}_\text{[${C}$] short-range activation} \biggr] \nonumber \\
&\hskip0.3cm \text{and}  \hskip0.3cm \underbrace{\sum_{i=1}^{N_\text{X}}{ \mathds{1}}_{\Omega^{\textbf{z}}_{\text{podia}}} (\textbf{X}_i) > \beta \sum_{i=1}^{N_\text{M}}{ \mathds{1}}_{\Omega^{\textbf{z}}_{\text{podia}}}(\textbf{M}_i)}_{\text{[${D}$] long-range inhibition}}  \hskip0.3cm \text{and} \hskip0.3cm  \underbrace{\sum_{i=1}^{N_\text{X}}{ \mathds{1}}_{\Omega^{\textbf{z}}_{\text{crowd}}} (\textbf{X}_i) + \sum_{i=1}^{N_\text{M}}{ \mathds{1}}_{\Omega^{\textbf{z}}_{\text{crowd}}}(\textbf{M}_i) < \eta}_\text{[${E}$] condition to prevent overcrowding}. \label{eq:mbirth}
\end{align}
The parameters $\phi, \psi, \alpha$, and $\beta$ in Eqns. (\ref{eq:xbirth}--\ref{eq:mbirth}) are involved in short-range activation and long-range inhibition; $\kappa$ and $\eta$ prevent overcrowding; and $d$ is a switch parameter that allows us to select whether or not to include Mechanism II in a given simulation. We note that $\phi, \psi, \alpha, \beta, \kappa$, and $\nu$ take on the same (or slightly altered) values as in the body model \cite{volkening} (see Table \ref{table:birth}). When we alter these parameters, we do so to reduce long-range inhibition and increase short-range activation in cell birth; additionally, we replace $\eta$ and $\kappa$ (involved in cell overcrowding) with the values used in the more recent model \cite{volkening2}.

Following the example in \cite{volkening}, we also include a small amount of random cell birth in some simulations. When included, we model random differentiation in exactly the same way as in the body model \cite{volkening}. For each of the $N^\text{M}_\text{diff}(t)$ locations $\textbf{z}$ for possible $M$ birth at day $t$, if there are no cells in $\Omega^\textbf{z}_\text{rand}$ (a disk of radius $100$ $\mu$m centered at $\textbf{z}$), we place a new $M$ cell at $\textbf{z}$ with probability $p_\text{M} \Delta t_\text{mig,birth}$. The corresponding rule also applies for $X$ cell birth, with probability $p_\text{X}\Delta t_\text{mig,birth}$, as in the body model \cite{volkening}.

The conditions in Eqns. (\ref{eq:xbirth}--\ref{eq:mbirth}) can be expressed as 
\begin{align*} \left[ \hskip0.1cm \left( \hskip0.1cm  A  \hskip0.1cm  \&  \hskip0.1cm  B  \hskip0.1cm  \right)  \hskip0.1cm  ||  \hskip0.1cm  C   \hskip0.1cm \right]  \hskip0.1cm \&  \hskip0.1cm D  \hskip0.1cm  \&   \hskip0.1cm E.
\end{align*}
Adjusting $d$ in Eqns. (\ref{eq:xbirth}--\ref{eq:mbirth}) then leads to two different dynamics:
 \begin{itemize}[noitemsep,nolistsep]
 \item \emph{Control case (body-model birth):} When $d = -1$ $\mu$m, Eqns. (\ref{eq:xbirth}--\ref{eq:mbirth}) reduce to $ C  \hskip0.1cm  \&   \hskip0.1cm D \hskip0.1cm \& \hskip0.1cm E $, the base rules of short-range activation and long-range inhibition (with overcrowding prevented) used in \cite{volkening}. 
 \item \emph{Mechanism II:} When $d = 150$ $\mu$m in Eqns. (\ref{eq:xbirth}--\ref{eq:mbirth}), we allow cells to differentiate in a small neighborhood of the proximal edge of the fin (when cell density is low) according to long-range activation alone (see Fig.~\ref{model}g). This models cells entering the fin from the proximal edge, where the fin connects with the body. 
 \end{itemize}

\subsubsection{Cell competition} \label{death}

After cell migration and birth, we evaluate every cell for possible death once on each simulated day of development. Motivated by ablation experiments on the fish body \cite{Nakamasu}, we specify short-range competition between $M$ and $X$. We also include a long-range survival signal from $X$ to $M$ cells: black cells die with small probability informed by \cite{Nakamasu} when there are insufficient $X$ cells present in adjacent interstripes. Our rules are the same as those used in the body model \cite{volkening}, and we reproduce them below:
\begin{align} &\underbrace{\sum_{i=1}^{N_\text{X}}{ \mathds{1}}_{\Omega^{\textbf{M}_j}_{\text{loc}}} \textbf{X}_i > \mu \sum_{i=1}^{N_\text{M}}{ \mathds{1}}_{\Omega^{\textbf{M}_j}_{\text{loc}}}\textbf{M}_i}_\text{ local competition with $X$} \hskip0.3cm \text{or} \hskip0.3cm \underbrace{P_i\sum_{i=1}^{N_\text{M}}{ \mathds{1}}_{\Omega^{\textbf{M}_j}_{\text{podia}}} \textbf{M}_i > \xi \sum_{i=1}^{N_\text{X}}{ \mathds{1}}_{\Omega^{\textbf{M}_j}_{\text{podia}}}\textbf{X}_i}_\text{long-range survival signals from $X$} \hskip0.3cm \Longrightarrow  \hskip0.3cm \text{death of $M$ cell at $\textbf{M}_j$,} \label{eq:mdeath} \\
&\underbrace{ \sum_{i=1}^{N_\text{M}}{ \mathds{1}}_{\Omega^{\textbf{X}_j}_{\text{loc}}} \textbf{M}_i > \nu \sum_{i=1}^{N_\text{X}}{ \mathds{1}}_{\Omega^{\textbf{X}_j}_{\text{loc}}}\textbf{X}_i }_\text{local competition with $M$} \hskip0.3cm \Longrightarrow \hskip0.3cm \text{death of $X$ cell at $\textbf{X}_j$},  \label{eq:xdeath}
\end{align}
where $P_i$ is a Bernoulli random variable with mean $p_\text{death}$ based on \cite{Nakamasu} and the parameters $\mu$, $\xi$, and $\nu$ are involved in short-range activation and long-range inhibition. We provide our cell death parameters in Table~\ref{table:death} in Appendix~\ref{appendix:parameters}. The values of $\nu$ and $p_\text{death}$ are the same as in \cite{volkening}. In some simulations, we adjust $\mu$ and $\xi$ from their body-model equivalents. When we adjust these two parameters, we do so to increase the strength of $M$ and $X$ competition locally. This helps developing stripes (respectively, interstripes) gradually extend from proximal to distal without being interrupted by the appearance of $X$ (respectively, $M$ cells) blocking their path.

\subsubsection{A note on empirical similarities on the fish body and tailfin}

We actively made the choice to base our fin model on the body model \cite{volkening} to test whether the same $M$ and $X$ interactions can account for patterning in both regions. Nevertheless, it is worth noting where the rules that we use also have a basis in the biological data for fins. First, the rules and parameters for cell birth in \cite{volkening} are heavily based on ablation experiments \cite{Nakamasu} that were performed on the fish body. Ablation severely damages the fins, so the same data is not available for the tailfin \cite{kondoTuringQuestion}. However, experiments with a temperature-sensitive mutation \cite{ParTur130} have indicated that $X$ support $M$ cells at long range on the fins \cite{kondoTuringQuestion}. Furthermore, \cite{goodrich1954analysis} provided early evidence for $M$ and $X$ competition, as well as a repulsion of $M$ by $X$ cells on the fins.

\section{\emph{In silico} pattern formation on the tailfin} \label{results}

By simulating the agent-based that we adapted from the body model \cite{volkening} under the five fin-specific mechanisms that we identified in Sec.~\ref{proposed}, we now explore how skin growth, bone rays, and the body--fin interface may affect patterning on the tailfin. Our goal is to help begin to reconcile the differences in pattern dynamics that are present on the fish body and tailfin. Our main result is a demonstration that two cell types can produce horizontal stripes on the caudal fin without iridophores. We find that there are several possible mechanisms that can align stripes, suggesting directions for future work related to pattern robustness.

In this initial modeling study of tailfin patterns, we adopt an exploratory perspective and focus on reproducing macro-scale features (with the exception of measurements in Fig.~\ref{Fig2}b and \ref{Fig3}b). It is important to note that the model \cite{volkening} on which we base our fin work agrees with several quantitative measurements on the fish body.  Future studies can fine-tune parameters and length scales to account for new data on fins. Based on our view of the timelines of wild-type development presented in \cite{Parichy}, we identify four broad qualitative measurements of model performance. We consider the model successful if (1) horizontal stripes develop; (2) stripes and interstripes form by extending from proximal to distal \cite{tu2010clonal,eom2015long}; (3) stripes form initially radially, but later adopt a horizontal alignment; and (4) four horizontal  stripes are fully formed (e.g., have roughly reached the distal edge of the fin) at stage J++ ($\approx 100$ dpf). We run many stochastic simulations and seek to meet these conditions consistently.

As the initial condition for our simulations, we start with a pattern that approximates the tailfin at $18$ dpf (see Appendix~\ref{appendix:initial}). We specify a central $M$ strip and a random distribution of $X$ cells (e.g., see Fig.\ \ref{Fig1}a). We do not include $M$ randomly across the domain, though black cells can be seen dispersed on the fin at the PB stage in images in \cite{Parichy}. As we discuss in Sect.~\ref{discussion}, we assume that these cells are in an earlier, larval form. The works \cite{tu2010clonal, rawls2000zebrafish}, for example, mention the presence of several types of $M$ cells on the fin.

\subsection{Mechanism I: Distal epithelial growth}

\begin{figure*}[t]\centering
\includegraphics[width=0.8\textwidth]{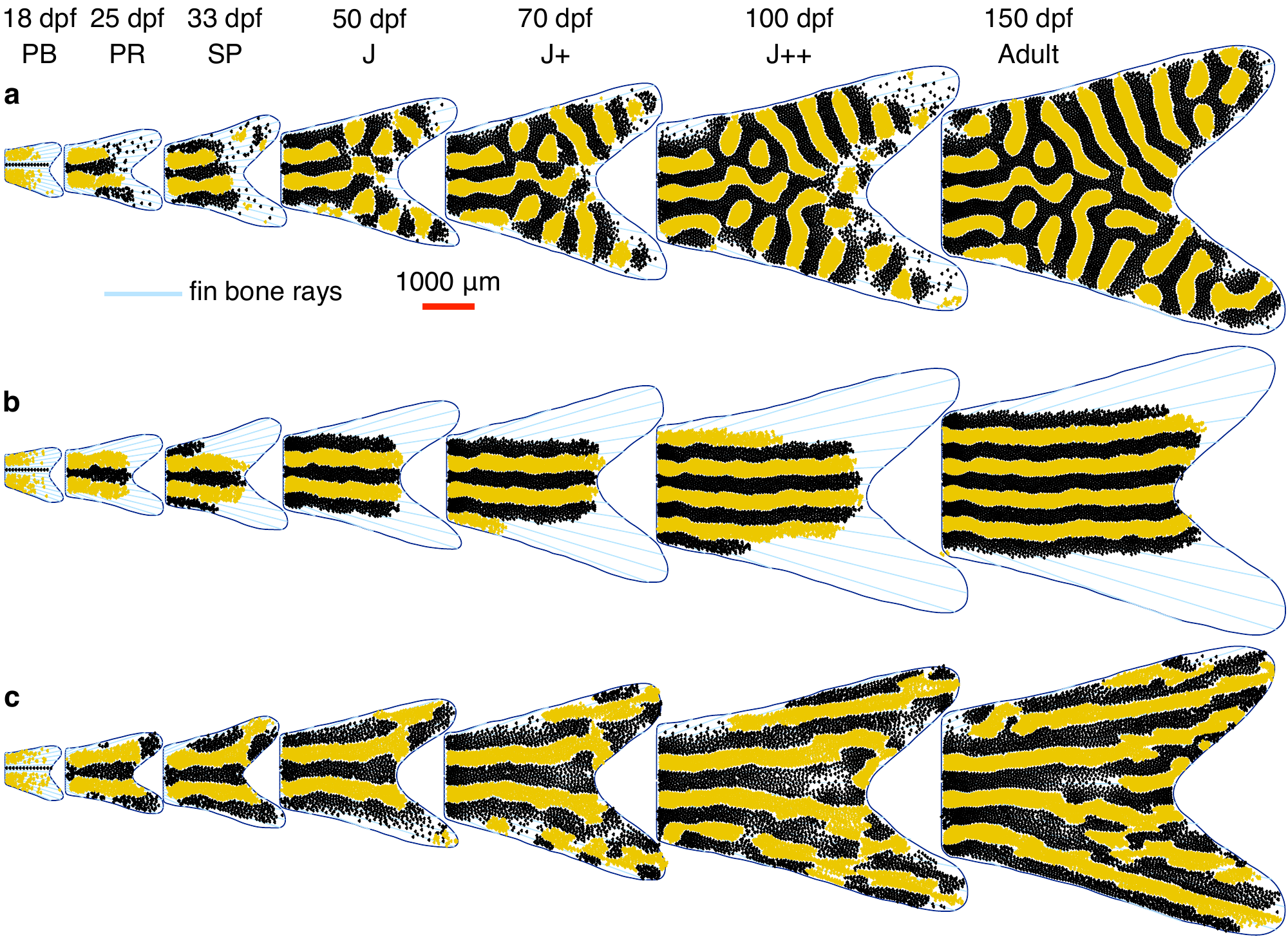} 
\caption{\label{Fig1} Simulated pattern formation under Mechanisms~I--III. (a) Under Mechanism I, our cell interactions (and the associated parameters) are the same as in the body model \cite{volkening}. We assume skin growth occurs at the distal edge of the fin, so that cell positions are not altered due to domain growth (see Fig.~\ref{model}a). In this case, random differentiation disrupts pattern alignment. (b) Under Mechanisms I \& II, we no longer allow random differentiation, and we instead model cells as interacting with the developing body pattern or precursors at the body--fin interface (see Fig.~\ref{model}g). To ensure pattern formation keeps pace with growth in this setting, we reduce $\Delta t_\text{mig, birth}$, allowing for more frequent cell birth. Notably, Mechanisms I \& II together produce patterns that spread from proximal to distal, but the stripes never appear radially as in Fig.~\ref{ablCartoon}d. (c) Under uniform epithelial growth with body-alignment cues (Mechanism II \& III), we stretch cell positions radially as the fin grows. This mechanism is able to produce aligned stripes, but stripe--interstripe boundaries are tortuous. The red scale bar is $1000$ $\mu$m and applies to all of the fins in this figure}
\end{figure*}

As an initial test of our model, we simulate patterning under the assumption of distal epithelial growth, so that cell positions remain unaltered during fin growth. Our rules for cell behavior in this setting are the same as those implemented in the body model \cite{volkening} (e.g., we set the switch parameters in Eqn.~(\ref{eq:m}) and Eqns.~(\ref{eq:xbirth}--\ref{eq:mbirth}) to $\zeta = -1$ cells and $d = -1$ $\mu$m, respectively). As we show in Fig. \ref{Fig1}a, epithelial growth at the distal end of the fin does not account for stripe directionality under the body model \cite{volkening}. The strip of $M$ cells in our initial condition at $18$ dpf is sufficient to specify stripe alignment proximally, but the dorsal pattern lacks horizontal directionality. This loss of horizontal alignment is not surprising when compared to the ablation experiments \cite{Yamaguchi} simulated in \cite{volkening}. In the absence of a pre-pattern in a portion of the domain (whether due to ablation on the fish body or empty space in the fin lobes), stripes form with lost directionality but maintained width.

The fact that the stripes continue to form horizontally near the proximal edge of the fin in Fig.~\ref{Fig1}a is most likely due to the strip of $M$ cells present in our initial condition in this region. We expect that the horizontal alignment that we observe there is also related to the smaller width of the fin near the proximal edge. Because we only allow random birth at low cell densities, smaller domains reduce the space available for random differentiation to operate and disrupt patterning. The appearance of vertical stripes within the fin lobes is particularly interesting and suggests avenues for future work, as several studies (e.g., \cite{ParTur130}) have observed vertical stripes in fin experiments.

\subsection{Mechanisms I \& II: Distal epithelial growth with alignment cues from the body}

In Fig.~\ref{Fig1}a, cell differentiation at random locations in the lobes seems to disrupt stripe alignment. We therefore do not include random birth here (e.g., we set $p_\text{M} = p_\text{X} = 0$). Instead, we include special dynamics in a narrow region near the body--fin interface: we model alignment cues from the body (Mechanism II) by setting $d = 150$ $\mu$m in Eqns.~(\ref{eq:xbirth}--\ref{eq:mbirth}). Near the body--fin interface, we then allow cells to be born based on long-range inhibition alone (see Fig.~\ref{model}g).  It is important to note that we cannot remove random birth if we do not implement Mechanism II: without either random birth or Mechanism II, there would be no new stripes added to our initial condition, as there would be no like cells present locally to activate the new formation of stripes or interstripes (see the form of our conditions in Eqns.~(\ref{eq:xbirth}--\ref{eq:mbirth})).

Our special condition for cell birth near the body--fin interface (Mechanism II) can be interpreted biologically in two main ways: first, precursors at the body--fin interface could conceivably interact with the developing body pattern. In particular, allowing cells to appear entirely due to long-range signals (without the need for established cells nearby) could model cells differentiating from precursors or existing cells at the body--fin interface. Second, Mechanism II can be viewed as modeling cells entering the fin through migration from the body pattern.

As we show in Fig.~\ref{Fig1}b, combining Mechanisms I and II produces straight stripes that form by spreading from proximal to distal across the tailfin. In order to obtain good performance under these mechanisms, we find that it is critical that our time step for cell migration and birth ($\Delta t_\text{mig,birth}$) is small and that the number of locations evaluated for $M$ birth is sufficiently large. This ensures that pattern formation keeps pace with fin growth, and it allows stripes to form by the highly controlled addition of pigment cells to the developing patterns. We obtain patterns that are similar to those in Fig.~\ref{Fig1}b in roughly $65$\% of $40$ stochastic simulations. The most frequent altered patterns feature straight stripes that form at a slight angle away from the horizontal, suggesting that it is not sufficient to rely on highly controlled cell birth to specify horizontal pattern alignment robustly.

\subsection{Mechanisms II \& III: Uniform epithelial growth with alignment cues from the body}

Under Mechanism III, we assume that the growing skin on the tailfin stretches across the bone rays, expanding everywhere as additional segments are added to the rays at their distal edges \cite{goldsmith2006developmental,goldsmith2003saltatory}. We implement this by scaling cell positions radially as the fin grows (see Fig.~\ref{finGrowth}b--c). As we show in Fig.~\ref{Fig1}c, uniform growth can account for some stripe directionality. Although our simulations under this mechanism produce aligned stripes, they maintain a strongly radial structure throughout fin development. 

It is is important to note that, in addition to uniform epithelial growth, we also include both random birth and Mechanism II (alignment cues from the body pattern) in Fig.~\ref{Fig1}c. If we remove Mechanism II and implement only uniform growth with random birth, we obtain the same qualitative pattern behavior in Fig.~\ref{Fig1}c (results not shown). In both cases patterns form very quickly across the fin and maintain a radial nature. We suggest that one could fine-tune our parameters (e.g., particularly those related to cell birth rates) in the future to determine if uniform epithelial growth can produce more gradual pattern formation across the caudal fin.

\subsection{Mechanisms I, III, \& IV: $M$ migration along bone rays with distal epithelial growth and alignment cues from the body}

Empirical images in \cite{Parichy} suggest that $M$ cells appear in close association with the rays. Because the caudal fin is thin in comparison to the fish body, we suggest that it is plausible that the bones on the tailfin physically hamper $M$ movement across rays (Mechanism IV in Sect.~\ref{proposed}). For example, in an early study \cite{goodrich1954analysis} of the anal fin, $M$ cells were described as invading yellow regions generally along the bone rays. The study \cite{tu2010clonal} also noted that melanophore clones were distributed parallel to rays on the caudal fin. We implement Mechanism IV by setting the switch parameter $\xi$ to $1000$ cells in Eqn.~(\ref{eq:m}), so that the velocity of each $M$ cell is projected along its nearest bone ray (also see Fig.~\ref{model}e).

\begin{figure*}[t]\centering
\includegraphics[width=0.8\textwidth]{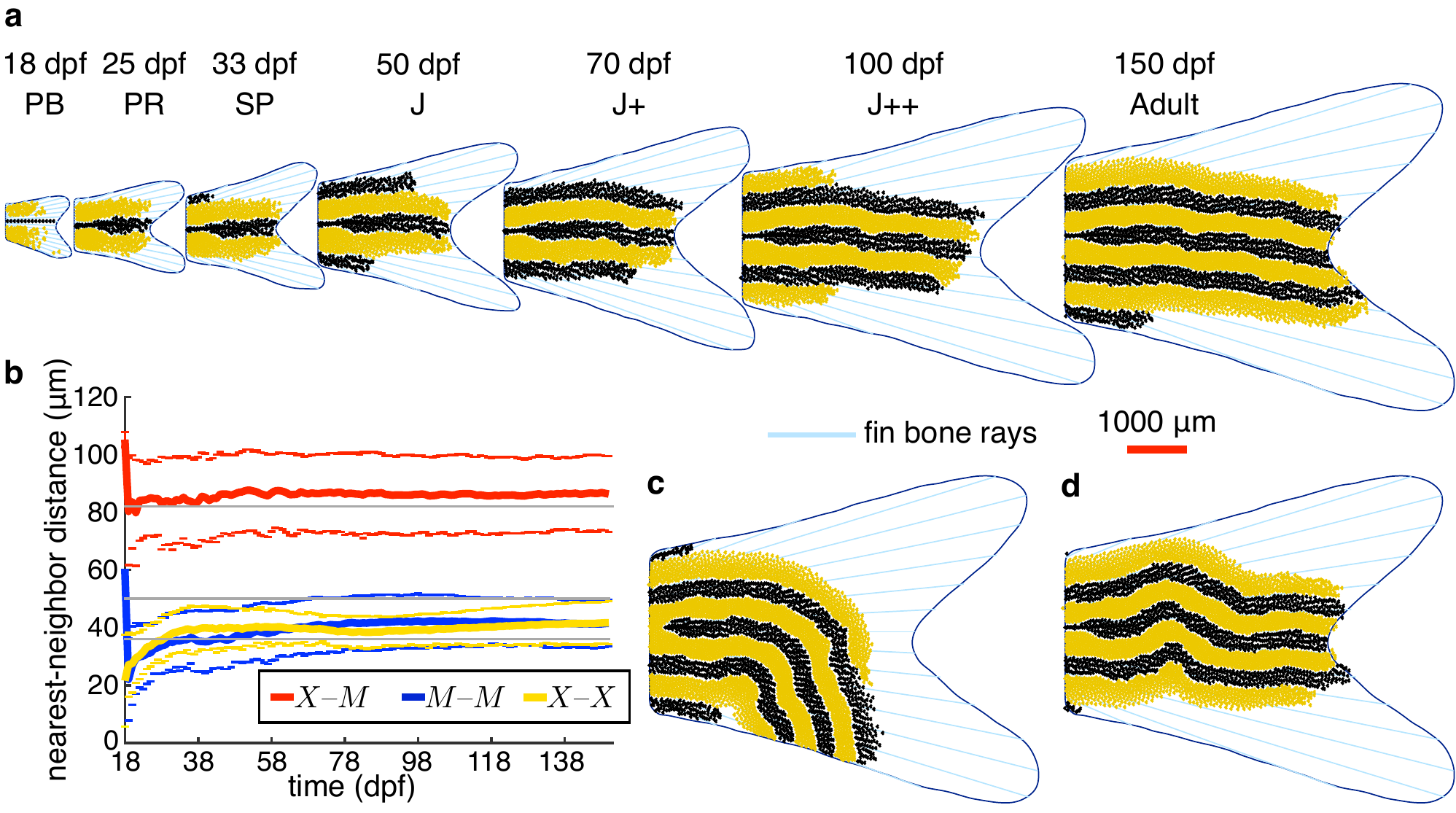}
\caption{\label{Fig2} Examples of simulated pattern formation under the combination of Mechanisms I, III, \& IV. (a) Based on early descriptions in \cite{goodrich1954analysis} and our observations of images in \cite{Parichy}, bone rays may affect $M$ movement. We implement this by setting $\xi = 1000$ cells in Eqn.~(\ref{eq:m}). Here we show an example of successful pattern formation under Mechanism IV (with distal epithelial growth and alignment cues from the body). We find that projecting $M$ migration along the bones preserves horizontal stripe alignment in roughly $60\%$ of $50$ stochastic simulations. (b) The mean nearest-neighbor distances between $M$ cells, between $X$ cells, and between $M$ and $X$ cells at stripe--interstripe boundaries for the simulation in (a) are similar to empirical measurements (see Appendix~\ref{appendix:figure} for how we calculate these measurements). In grey, we show empirical measurements from \cite{TakahashiMelDisperse}: this study reports that average $M$--$M$ distances are $50$~$\mu$m, $X$--$X$ distances are $36$ $\mu$m, and $X$--$M$ distances at stripe--interstripe boundaries are $82$ $\mu$m. We expect that our $M$--$M$ distances are smaller because projecting $M$ movement along the rays may remove components of $M$ velocity due to repulsion from other $M$ cells. The dashed curves indicate standard deviation. The most frequent alterations to our target patterns include (c) severely curved or bent stripes and (d) wavy stripes (both shown at $150$ dpf). The red scale bar is $1000$ $\mu$m and applies to all of the fins in this figure
}
\end{figure*}

As we show in Fig.~\ref{Fig2}a, if we enforce $M$ migration along the rays, our model is able to generate stripe patterns with horizontal alignment. These patterns form by gradually extending distally, and three dark stripes reach the distal end of the fin domain around $100$ dpf. Nevertheless, Mechanism IV (together with distal epithelial growth and alignment cues from the body) is not reliable across our stochastic simulations. Based on a visual classification of $50$ simulations, we find that roughly $40$\% of our simulations do not produce horizontal patterns under the assumption of $M$ migration along the rays. The most frequent altered patterns include bent, curved, and wavy stripes (see Fig.~\ref{Fig2}c--d).

\subsection{Mechanisms I, III, and V: $M$ birth in association with bone rays with distal epithelial growth and alignment cues from the body}\label{bonebirth}

\begin{figure*}[t]\centering
\includegraphics[width=0.8\textwidth]{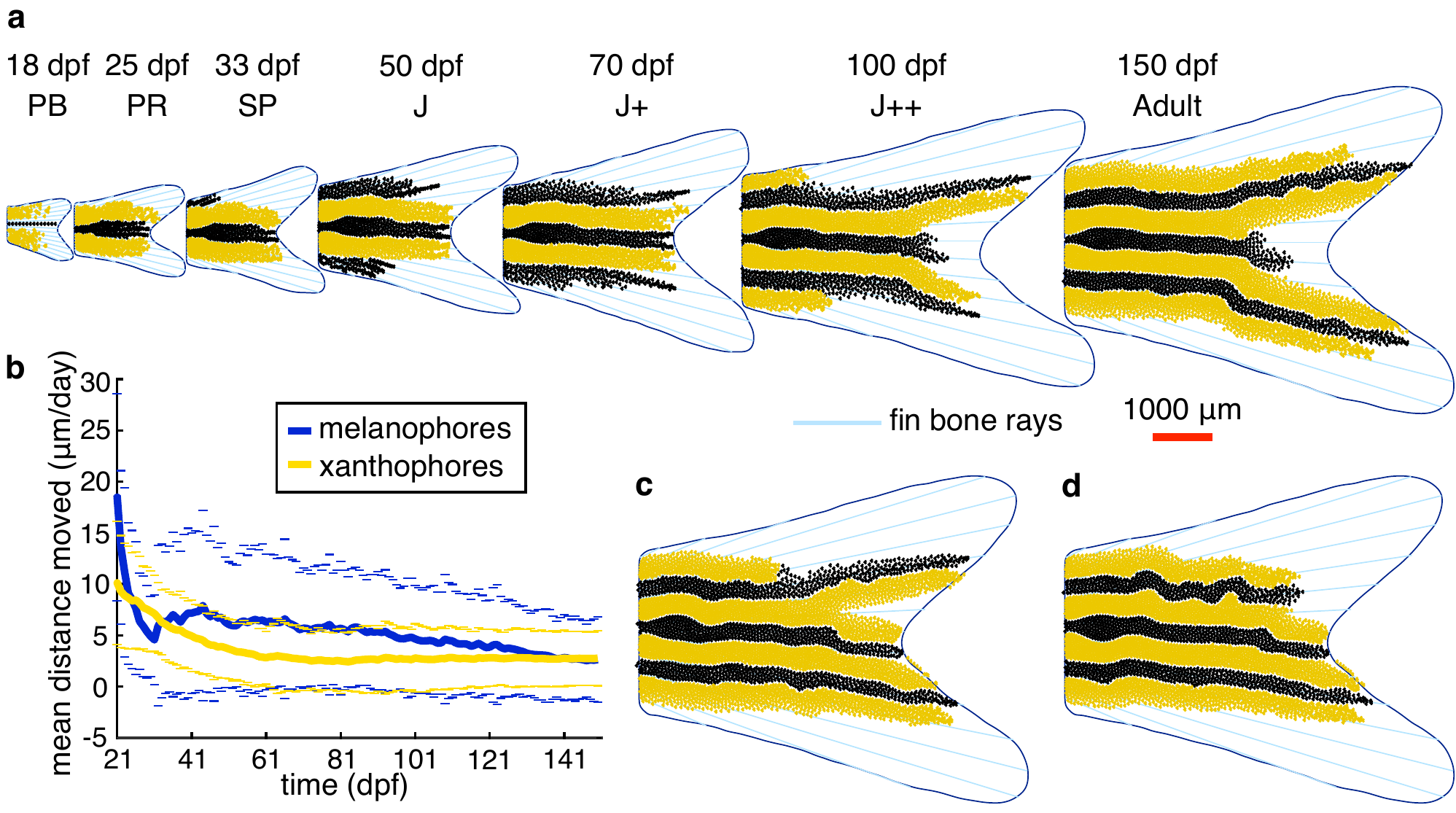}
\caption{\label{Fig3} Simulated pattern formation under the combination of Mechanisms I, III, \& V. (a) Based on our observations of images in \cite{Parichy}, as well as early notes about anal fin patterns in \cite{goodrich1954analysis}, $M$ cells seem to appear in close association with bone rays. One explanation (tested in Fig.~\ref{Fig2}) for this association is that $M$ cells migrate along bone rays. Alternatively, it is possible that $M$ cells differentiate from precursors associated with the rays (Mechanism V). We find that patterns robustly form by extending distally across the fin under Mechanism V (with distal epithelial growth and alignment cues from the body). (b) The mean distance moved by $M$ and $X$ cells per day is similar to \emph{in vivo} \cite{TakahashiMelDisperse} and \emph{in silico} \cite{volkening} measurements of cell speed. The thick curves indicate the mean distances moved across all the cells on the domain (for the simulation that we present in (a)). The dashed curves indicate standard deviation. See Appendix~\ref{appendix:figure} for additional details. (c) Under Mechanism V, patterns often develop with a branched ``Y'' gap in interstripes or stripes, but we also observe (d) some wavy stripe patterns without such gaps. Patterns in (c) and (d) are at $150$ dpf, and the red scale bar ($1000$ $\mu$m) applies to all of the fins in this figure}
\end{figure*}

As an alternative means of explaining our observations of $M$ cells in close association with the bones in images of caudal fins from \cite{Parichy}, it is possible that $M$ cells arise from precursors or other sources that are distributed along or between the rays. (As we mentioned earlier, the study \cite{tu2010clonal} noted that boundaries of melanophore-clone distributions are parallel to the bones on the tailfin.) To test this hypothesis, we select locations to evaluate for possible $M$ birth along the rays (see Appendix~\ref{appendix:cellBirth} for details). As we show in Fig.~\ref{Fig3}a, requiring that $M$ cells arise along the bones produces patterns that emerge from proximal to distal. Moreover, Mechanism V (with distal epithelial growth and alignment cues from the body) performs robustly across our stochastic simulations, and we show the final patterns that result from a few other representative simulations under this combination of mechanisms in Fig.~\ref{Fig3}c--d. Although our focus is on qualitative measurements in this first fin model, we also show the mean speeds of $M$ and $X$ cells in time (for the simulation in Fig.~\ref{Fig3}a) in Fig.~\ref{Fig3}b. These measurements are similar to empirical estimates \cite{TakahashiMelDisperse} that $M$ cells move roughly $80$--$100$ $\mu$m per week \emph{in vivo} (e.g., about $11$--$14$ $\mu$m per day) when they are active (prior to forming densely-packed stripes).

Although the patterns in Fig.~\ref{Fig3} form robustly, the stripes that develop under Mechanism V are strongly directed by the radial nature of the bones. This often results in stripes and interstripes that diverge in a ``Y'' shape at the fin cleft. We suggest that mechanisms of $M$ birth may be different in different regions of the fin, as requiring that melanophores are born in association with the bones across the full tailfin in our model produces patterns that too strongly adhere to the radial geometry of the rays.

\section{Discussion}\label{discussion}

In recent years, empirical views of zebrafish skin patterning have changed significantly, as it has become clear that three main types of pigment cells, rather than two, are necessary for pattern development \cite{Frohnhofer,Singh,PatPLos}. When iridophores fail to emerge in \emph{shady} mutants \cite{Frohnhofer,Lopes}, dark spots form on the body, but stripe patterns remain intact on the tailfin. Moreover, on the body, the horizontal myoseptum provides an initial source of stripe alignment \cite{Frohnhofer}, but this marker is not present on the caudal fin. To begin to understand these differences between body and fin patterns, we presented an exploratory modeling study of wild-type stripe formation on zebrafish tailfins. We based our agent-based model for melanophore and xanthophore interactions on the prior body model \cite{volkening}, and we traced images of zebrafish from \cite{Parichy} to produce growing caudal-fin domains. 

We highlighted five different patterning mechanisms that could be at work on the tailfin, and we showed that  several of these mechanisms are capable of producing stripes. We find that uniform skin growth can account for radial stripe alignment by stretching pigment cell positions on the fish skin. Under distal epithelial growth, if we incorporate some alignment cues from the developing body stripes into our model indirectly, stripe patterns can form horizontally by progressing from proximal to distal. In our model, pattern robustness under distal epithelial growth depends on melanophore birth keeping pace with fin growth or on bone rays controlling $M$ cell birth or migration. Possibly related, \cite{tu2010clonal} showed that the boundaries of melanophore-clone distributions are parallel to the rays on the tailfin. 

Most importantly, our work supports observations related to the \emph{shady} mutant \cite{Frohnhofer}: we find that the interactions of melanophores and xanthophores (without iridophores) can indeed lead to wild-type patterns on the tailfin. We suspect, however, that these cell interactions benefit from external cues (in the form of skin growth, bone rays, the developing body pattern, or other environmental cues), as horizontal-stripe alignment is not robust across our stochastic simulations. We find the most robust results when we assume melanophores arise from precursors associated with the bone rays, but this mechanism also leads to stripe patterns that adhere too closely to the radial nature of the bones. We therefore hypothesize that cues present in the cellular environment may be different in different regions of the fin. For example, it is possible that a cellular signal diffuses from the body--fin interface, that the thickness of the fin affects cell behavior, or that pigment cells are different in different regions of the fin \cite{tu2010clonal, rawls2000zebrafish}. As an initial means of accounting for such differences, one could set the switch parameters in Eqns.~(\ref{eq:m}) and (\ref{eq:xbirth}--\ref{eq:mbirth}) to more moderate values.

Our work provides an initial modeling frame of reference for fin patterning on zebrafish and suggests many places for improvement and extension in future model generations. For example, we focused largely on qualitative measurements, with a particular focus on obtaining horizontal stripe alignment and reproducing proximal-to-distal pattern development. As an additional qualitative measurement, one could seek to reproduce the appearance of random melanophores and xanthophores intermingled in the fin lobes (see images in \cite{Parichy} and the supplementary material in \cite{eom2015long}). However, the short-range melanophore--xanthophore competition in our model does not support this intermingling of unlike cells. We suspect that the cells present in the fin lobes are in an early form, and therefore are beyond the scope of a two-cell model. The works \cite{tu2010clonal, rawls2000zebrafish}, for example, mention the presence of several types of melanophores on the fin. In the future, it would be interesting to study how cell stage, age, or lineage impact pattern formation. Finally, comparing quantitative measurements on cell behavior with fin data would be useful in future models, as would large-scale, quantitative studies of \emph{in vivo} and \emph{in silico} patterns to more carefully explore pattern robustness and variability.

\begin{center}Fin.\end{center}

\appendix
\FloatBarrier

\section{Appendix: Parameters} \label{appendix:parameters}

With the exception of the two switch parameters (namely, $\zeta$ in Eqn.~(\ref{eq:m}) and $d$ in Eqns.~(\ref{eq:xbirth}--\ref{eq:mbirth})) that we use to implement Mechanism II and IV, we summarize all of the parameters involved in our rules for cell migration, birth, and death in Tables~\ref{table:mig}, \ref{table:birth}, and \ref{table:death}, respectively. We give the simulation-specific values of the switch parameters by figure in Appendix~\ref{appendix:figure}. We set $d = 150$ $\mu$m when we include Mechanism II, and we set $\zeta = 1000$ cells when we test Mechanism IV.

     \begin{table}[h]
     \centering
     \begin{tabular}{l  l  l  l  }
         {\bf{Name}} & {\bf{Body-model value}} & {\bf{Fin-specific value}} & {\bf{Description}}  \\ \toprule 
         $\Delta t_\text{mig}$ & $1$ day & $1/3$ day & Numerical time step for cell migration \\ \midrule
     $R^\text{MM}$ & $62$ $\mu$m/day & 24.8 $\mu$m/day & Melanophore-on-melanophore repulsion \\ \midrule
          $R^\text{XX}$ & 50 $\mu$m/day & 20 $\mu$m/day  & Xanthophore-on-xanthophore repulsion\\ \midrule
     $R^\text{XM}$ & 137 $\mu$m/day  & 35 $\mu$m/day& Xanthophore-on-melanophore repulsion\\ \midrule
          $R^\text{MX}$ & 113 $\mu$m/day & 30 $\mu$m/day & Melanophore-on-xanthophore repulsion \\ \midrule
          $A^\text{MX}$ & 163 $\mu$m/day & 0 $\mu$m/day & Melanophore-on-xanthophore attraction \\ \midrule
         $r_\text{MM}$ & 20 $\mu$m & 40 $\mu$m& Appears in $Q^\text{MM}$\\ \midrule
          $r_\text{XX}$ &11 $\mu$m  & 31 $\mu$m &Appears in $Q^\text{XX}$ \\ \midrule
     $r_\text{XM}$ & 20 $\mu$m & 40 $\mu$m&Appears in $Q^\text{XM}$\\ \midrule
          $r_\text{MX}$ & 20 $\mu$m & 40 $\mu$m &Appears in $Q^\text{MX}$ \\ \midrule
        $a_\text{MX}$ &12 $\mu$m & 40 $\mu$m&Appears in $Q^\text{MX}$\\ \midrule
        $R^\text{bnd}$ & 137 $\mu$m/day & 50 $\mu$m/day & Repulsion from the fin boundary\\ \midrule
        $r_\text{bnd}$ & 20 $\mu$m & 20 $\mu$m& Appears in boundary force potential $Q^\text{bnd}$\\
     \bottomrule
\end{tabular}
\caption{{Summary of parameters for cell migration in Eqns. (\ref{eq:x}--\ref{eq:m}) and (\ref{eq:bound}). We note that the model \cite{volkening} reported $R^\text{MM} = 250$ $\mu$m, $R^\text{XX} = 200$ $\mu$m, $R^\text{XM} = 550$ $\mu$m, $R^\text{MX} = 450$ $\mu$m, $A^\text{MX} = 650$ $\mu$m, $R^\text{bnd} = 137$ $\mu$m, and $\Delta t_\text{mig} = 0.25$ days, but these parameters were off by a factor of four; the correct values used in \cite{volkening} are instead $R^\text{MM} = 62.5$ $\mu$m, $R^\text{XX} = 50$ $\mu$m, $R^\text{XM} = 137.5$ $\mu$m, $R^{MX} = 450$ $\mu$m, $A^\text{MX} = 162.5$ $\mu$m, $R^\text{bnd} = 137.5$ $\mu$m, and $\Delta t_\text{mig} = 1$ day. Here we have rounded the repulsion and attraction parameters down or up to the nearest whole number as indicated  \label{table:mig}}}
\end{table}

     \begin{table}[h]
     \centering
     \begin{tabular}{l  l  l   p{8.5cm} }
         {\bf{Name}} & {\bf{Body-model value}} & {\bf{Fin-specific value}} & {\bf{Description}}  \\ \toprule
              $\Delta t_\text{mig,birth}$ &$1$ day & $1/3$ day & Numerical time step for birth and migration\\ \midrule
     $n^\text{M}_{\text{diff}}$ &Varies  &$600$ locations & Initial condition for number of locations selected for possible $M$ differentiation per $\Delta t_\text{mig,birth}$ days\\ \midrule
          $n^\text{X}_{\text{diff}}$ &Varies  & $600$ locations &Initial condition for number of locations selected for possible $X$ differentiation per $\Delta t_\text{mig,birth}$ days\\ \midrule
          $d_{\text{loc}}$ &$75$ $\mu$m &$82$ $\mu$m & Radius of short-range interaction disk $\Omega_\text{loc}$ \\ \midrule
                             $d_{\text{crowd}}$ & $82$ $\mu$m  & $82$ $\mu$m & Radius of overcrowding disk $\Omega_\text{crowd}$ \\ \midrule
                        $d_{\text{rand}}$ &$100$ $\mu$m &$100$ $\mu$m & Radius of disk for random birth $\Omega_\text{rand}$ \\ \midrule
$d_{\text{podia}}$ &$318$ $\mu$m &$318$ $\mu$m & Inner radius of long-range annulus $\Omega_\text{podia}$ \\ \midrule
$w_{\text{podia}}$ &$25$ $\mu$m &$25$ $\mu$m & Width of long-range annulus $\Omega_\text{podia}$ \\ \midrule
        $\alpha$ & $1$& $0.5$ &Lower bound on $\frac{\sum_{i=1}^{N_\text{M}}\textbf{M}_i}{\sum_{i=1}^{N_\text{X}}\textbf{X}_i}|_{\Omega_{\text{loc}}}$ for $M$ birth \\ \midrule
    $\beta$ &$3.5$ & $2.5$ & Lower bound on $\frac{\sum_{i=1}^{N_\text{X}}\textbf{X}_i}{\sum_{i=1}^{N_\text{M}}\textbf{M}_i}|_{\Omega_{\text{podia}}}$ for $M$ birth \\ \midrule
         $\eta$ &$6$ & $4$ & Upper bound on $(\sum_{i=1}^{N_\text{M}}\textbf{M}_i +\sum_{i=1}^{N_\text{X}}\textbf{X}_i)|_{\Omega_{\text{crowd}}}$ for $M$ birth \\ \midrule
          $\phi$ &$1.3$ & $1.3$ &Lower bound on $\frac{\sum_{i=1}^{N_\text{X}}\textbf{X}_i}{\sum_{i=1}^{N_\text{M}}\textbf{M}_i}|_{\Omega_{\text{loc}}}$ for $X$ birth \\ \midrule
    $\psi$ &$1.2$ & $1$ & Lower bound on $\frac{\sum_{i=1}^{N_\text{M}}\textbf{M}_i}{\sum_{i=1}^{N_\text{X}}\textbf{X}_i}|_{\Omega_{\text{podia}}}$ for $X$ birth\\ \midrule
     $\kappa$ & $10$& $6$ & Upper bound on $(\sum_{i=1}^{N_\text{M}}\textbf{M}_i +\sum_{i=1}^{N_\text{X}}\textbf{X}_i|_{\Omega_{\text{crowd}}}$ for $X$ birth \\ \midrule
          $p_{\text{M}}$ & $0.03$ & $0$ & Probability of random $M$ birth per day \\ \midrule
     $p_{\text{X}}$ & $0.005$ & $0$ & Probability of random $X$ birth per day\\ 
     \bottomrule
\end{tabular}
\caption{{Summary of our parameters for cell birth (see Eqns.~(\ref{eq:xbirth}--\ref{eq:mbirth})) and model length scales. In Appendix~\ref{appendix:figure}, we note whether we use the body-model or fin-specific values to produce each figure in the manuscript. Body-model values refer to the parameters used in the prior body model \cite{volkening}\label{table:birth}}}
\end{table}

     \begin{table}[h]
     \centering
     \begin{tabular}{l  l  l  l }
         {\bf{Name}} & {\bf{Body-model value}} & {\bf{Fin-specific value}} & {\bf{Description}}  \\ \toprule
              $\Delta t_\text{death}$ &$1$ day &$1$ day & Numerical time step for cell death\\ \midrule
             $\mu$ &$1$ &$2$ &Lower bound on $\frac{\sum_{i=1}^{N_\text{X}}\textbf{X}_i}{\sum_{i=1}^{N_\text{M}}\textbf{M}_i}|_{\Omega_{\text{loc}}}$ for $M$ death\\ \midrule
     $\nu$ &$1$ & $1$ &Lower bound on $\frac{\sum_{i=1}^{N_\text{M}}\textbf{M}_i}{\sum_{i=1}^{N_\text{X}}\textbf{X}_i}|_{\Omega_{\text{loc}}}$ for $X$ death \\ \midrule
       $\xi$ &$1.2$ & $1.7$ &Lower bound on $\frac{\sum_{i=1}^{N_\text{M}}\textbf{M}_i}{\sum_{i=1}^{N_\text{X}}\textbf{X}_i}|_{\Omega_{\text{podia}}}$ for $M$ death \\ \midrule
     $p_{\text{death}}$ &$0.0333$ & $0.0333$ & Probability of $M$ death per day due to long-range effects\\ 
     \bottomrule
\end{tabular}
\caption{{Summary of our parameters for cell death in Eqns.~(\ref{eq:mdeath}--\ref{eq:xdeath}). In Appendix~\ref{appendix:figure}, we note whether we use the body-model or fin-specific values to produce each figure in the manuscript. Body-model values refer to the parameter values used in \cite{volkening} to simulate patterning on the zebrafish body \label{table:death}}}
\end{table}

{\section{Appendix: Simulation conditions}\label{appendix:simulation}

We used MATLAB $9.3$, The MathWorks, Inc., Natick, MA, USA to simulate our model of cell interactions on growing fins. Our code is available from the corresponding author on request. We now describe our model implementation (Appendix~\ref{appendix:flow}), our boundary conditions (Appendix~\ref{appendix:boundary}), our initial conditions (Appendix~\ref{appendix:initial}), and our methods for selecting cell-birth locations in tailfin domains and implementing Mechanism V (Appendix~\ref{appendix:cellBirth}). In Appendix~\ref{appendix:figure}, we summarize the parameters and simulation conditions associated with our simulations in Figures \ref{Fig1}, \ref{Fig2}, and \ref{Fig3}.

\subsection{Model implementation}\label{appendix:flow}
Our simulation begins with an initial condition (see Appendix~\ref{appendix:initial}) at $t=18$ dpf. We update cell positions from day $t$ to day $t+1$ in eight steps:
\begin{enumerate}[noitemsep,nolistsep]
\item Set the cycle counter $c$ for migration and birth to $c=1$.
\item Perform one step $\Delta t_\text{mig,birth}$ of migration (e.g., solve Eqns.~(\ref{eq:x}) and (\ref{eq:v}) using the Euler forward scheme with time step $\Delta t_\text{mig,birth}$). All of our cells migrate simultaneously. We also specify repulsive forces from the discretized fin boundaries at each step $\Delta t_\text{mig,birth}$ of migration (see Appendix~\ref{appendix:boundary}).
\item Increment $N^\text{M}_\text{diff}(t)$ and $N^\text{X}_\text{diff}(t)$ by $20$ locations each, so that $N^\text{M}_\text{diff}(t) = n^\text{M}_\text{diff} + 20(t-18)$ locations and $N^\text{X}_\text{diff}(t) = n^\text{X}_\text{diff} + 20(t-18)$ locations.
\item Select $N^\text{M}_\text{diff}(t)$ and $N^\text{X}_\text{diff}(t)$ potential locations for $M$ and $X$ birth on the fin domain (outlined by our boundary curve at time $t+1$), respectively. Evaluate these locations (simultaneously) for cell birth based on Eqns.~(\ref{eq:xbirth}--\ref{eq:mbirth}) and random birth if included (e.g., if $p_\text{M} >0$ and $p_\text{X} >0$). Add the newly born cells to the domain at time $t+c\Delta t_\text{mig,birth}$.
\item Increment the cycle counter $c$ for migration and birth: $c= c+1$. 
\item If $c \Delta t_\text{mig,birth} = 1$ day, one day of migration and birth has been completed: go to Step 7. Otherwise, return to Step 2 with the cell positions at time $t+c\Delta t_\text{mig,birth}$.
\item Evaluate all of the cells for possible death (simultaneously) through Eqns.~(\ref{eq:mdeath}--\ref{eq:xdeath}). Remove any cells that have died from the domain at day $t+1$. We note that the time step for our cell death rules is always $\Delta t_\text{death} =1$ day.
\item If uniform domain growth is included, scale the cell positions at time $t+1$ using the fin domains and bone rays at times $t+1$ and $t+2$ days as we describe in Sect.~\ref{growth}. If distal epithelial growth is included, do not scale the cell positions. The result of this process is the updated cell positions at time $t+1$ days.
\end{enumerate}

\subsection{Boundary conditions}\label{appendix:boundary}

To help keep cells in the fin domain, we include wall-like boundary conditions. For each simulated day $t$, we discretize the associated fin boundary curve at time $t$ into $500$ points: $\{\textbf{F}_i(t)\}_{i=1,\ldots,500}$. We then specify repulsive forces from these boundary points to our cell agents at each time step of migration $\Delta t_\text{mig,birth}$:
\begin{align} \text{boundary force on $i$th cell at position $\textbf{C}_i$ } &=  -\sum_{j=1}^{500} \triangledown Q^\text{bnd}(\textbf{F}_j - \textbf{C}_i)  \label{bndpoint}
\end{align}
where $\textbf{C} \in \{\textbf{M}, \textbf{X}\}$ and 
\begin{align} Q^\text{bnd}(\textbf{d}) &=  R^\text{bnd} e^{-|\textbf{d}|/r_\text{bnd}}.\label{eq:bound}
\end{align}
These rules are an approximation of Neumann boundary conditions, and we note that a small number of cells escape from our domain in some simulations. Cells are more likely to escape  with increasing time, suggesting that it may be useful for future work to increase the number of boundary agents as the fin grows. When cells escape, we remove them from our final simulated images in post-processing.

\subsection{Initial condition}\label{appendix:initial}

The initial condition for our simulations is motivated by images in \cite{Parichy}. First, we specify a single horizontal strip of $M$ cells (separated $30$ $\mu$m apart) at the center of our fin domain (e.g., with $y$-coordinate $0$). Second, for $X$ cells, we consider a random distribution of cells that are concentrated more highly toward the proximal edge of the fin. We choose the $y$-coordinates of these positions by selecting $500$ points uniformly at random between the maximum and minimum $y$-coordinates for the discretized boundary curve that represents our initial domain at $18$ dpf (see Appendix~\ref{appendix:boundary}). We choose the $x$-coordinates for these points by taking the absolute value of $500$ points sampled from a normal distribution with mean $0$ and standard deviation $\sigma = 0.25 x_\text{max}(t)$, where $x_\text{max}(t)$ is the maximum $x$-coordinate for our discretized boundary curve at $18$ dpf.

As the penultimate step in setting our initial condition, we remove any $M$ or $X$ cells that fall within $25$ $\mu$m of our discretized boundary curve in Eqn.~\ref{eq:bound}. Finally, if more than $300$ of our selected $X$ locations fall in the domain, we use only the first $300$ such locations in our initial condition.

For the special case of Fig.~\ref{Fig1}c, after specifying our initial condition as above, we scale the cell positions to account for one day of uniform epithelial growth (for details on how we implement domain growth, see Sect.~\ref{growth}). We use these scaled cell positions as our initial condition for the simulations in Fig.~\ref{Fig1}c.

\subsection{Selecting cell birth locations }\label{appendix:cellBirth}

We consider two methods for selecting $N^\text{M}_\text{diff}(t)$ possible locations for $M$ birth:
\begin{itemize}[noitemsep,nolistsep]
\item \emph{Control case (similar to body-model birth):} We choose $2\times N^\text{M}_\text{diff}(t)$ locations uniformly at random in a rectangular region surrounding the fin, and we evaluate the first $N^\text{M}_\text{diff}(t)$ of these positions that are inside the fin domain for possible birth simultaneously.
\item \emph{Mechanism V:} Under Mechanism V, we first choose $N^\text{M}_\text{diff}(t)$ points uniformly at random from our discretized bone rays, namely $\{\textbf{B}_i^j\}$ in Eqn.~(\ref{bone}). For each such point $\textbf{B}_k$, we then choose a corresponding location to evaluate for $M$ birth by selecting a point uniformly at random in a ball of radius $r \sim \mathcal{N}(0, 2)$ $\mu$m around $\textbf{B}_k$. 
\end{itemize}
We always select potential $X$ birth locations in the same way as in the $M$ control case.

Lastly, prior to applying our rules for cell birth to the locations that we randomly selected as we outlined above, we require that these positions are strictly greater than $25$ $\mu$m away from our discretized boundary curve (see Eqn.~(\ref{bndpoint}) in Appendix~\ref{appendix:boundary}). If a randomly selected location is within $25$ $\mu$m of a discretized boundary point in $\{\textbf{F}_i(t)\}_{i=1,\ldots,500}$, we do not allow cell birth to occur at that location.

\subsection{Instructions for reproducing our figures} \label{appendix:figure}

We summarize the parameters for our simulated patterns in Figures \ref{Fig1}, \ref{Fig2}, and \ref{Fig3} below:
\begin{itemize}
\item Fig.~\ref{Fig1}a, Mechanism I (distal epithelial growth)
\begin{itemize}
\item Cell migration: We use the body-model values in Table~\ref{table:mig}.
\item Cell birth: We use the body-model values in Table~\ref{table:birth} with $n^\text{M}_\text{diff}= n^\text{X}_\text{diff} = 600$ locations.
\item Cell death: We use the body-model values in Table~\ref{table:death}.
\item Skin growth: We do not scale cell positions with domain growth.
\item Switch parameters: In Eqn.~(\ref{eq:m}), we use $\zeta = -1$ cells. In Eqns.~(\ref{eq:xbirth}--\ref{eq:mbirth}), we use $d = -1$ $\mu$m.
\end{itemize}
\item Fig.~\ref{Fig1}b, Mechanisms I and III (distal epithelial growth with alignment cues from the body)
\begin{itemize}
\item Cell migration: We use the body-model values in Table~\ref{table:mig} with one exception: $\Delta t_\text{mig,birth} = 0.25$ days (more frequent cell birth and migration than we use in Fig.~\ref{Fig1}a). 
\item Cell birth: We use the body-model values in Table~\ref{table:birth} with three exceptions: $\Delta t_\text{mig,birth} = 0.25$ days, $p_\text{M}  = 0$, and $p_\text{X} = 0$ (no random birth). We use $n^\text{M}_\text{diff}= n^\text{X}_\text{diff} = 600$ locations.
\item Cell death: We use the the body-model values in Table~\ref{table:death}.
\item Skin growth: We do not scale cell positions with domain growth.
\item Switch parameters: In Eqn.~(\ref{eq:m}), we use $\zeta = -1$ cells. In Eqns.~(\ref{eq:xbirth}--\ref{eq:mbirth}), we use $d = 150$ $\mu$m (special body--fin interface dynamics; see Fig.~\ref{model}g).
\end{itemize}
\item Fig.~\ref{Fig1}c, Mechanisms II and III (uniform epithelial growth with alignment cues from the body)
\begin{itemize}
\item Cell migration: We use the body-model values in Table~\ref{table:mig}. 
\item Cell birth: We use the body-model values in Table~\ref{table:birth} with $n^\text{M}_\text{diff}= 1200$ and $n^\text{X}_\text{diff} =600$ locations.
\item Cell death: We use the body-model values in Table~\ref{table:death}.
\item Skin growth: We scale cell positions along the bone rays with fin growth as we describe in Sect.~\ref{growth}.
\item Switch parameters: In Eqn.~(\ref{eq:m}), we use $\zeta = -1$ cells. In Eqns.~(\ref{eq:xbirth}--\ref{eq:mbirth}), we use $d = 150$ $\mu$m.
\end{itemize}
\item Fig.~\ref{Fig2}, Mechanisms I, III, and IV ($M$ migration along bone rays with alignment cues from the body and distal epithelial growth)
\begin{itemize}
\item Cell migration: We use the fin-specific values in Table~\ref{table:mig}.
\item Cell birth: We use the fin-specific values in Table~\ref{table:birth}.
\item Cell death: We use the fin-specific values in Table~\ref{table:death}.
\item Skin growth: We do not scale cell positions with domain growth.
\item Switch parameters: In Eqn.~(\ref{eq:m}), we use $\zeta =1000$ cells (so that $M$ movement is always projected along the bones). In Eqns.~(\ref{eq:xbirth}--\ref{eq:mbirth}), we use $d = 150$ $\mu$m.
\item Additional note on Fig.~\ref{Fig2}b: When we calculate the nearest-neighbor distances between cells, we only consider $M$--$M$ and $X$--$X$ distances that are less than $200$ $\mu$m (this ensures that cells that have escaped our fin domain or appear at low density do not affect our measurements of cell--cell distances in developing stripes). When we calculate $M$--$X$ distances at stripe--interstripe boundaries, we only consider measurements that are less than $110$ $\mu$m. We made this choice so that our $M$--$X$ distances measure stripe--interstripe separation (in comparison, \cite{TakahashiMelDisperse} showed that $M$ and $X$ cells are roughly $82$ $\mu$m apart at stripe--interstripe boundaries).
\end{itemize}
\item Fig.~\ref{Fig3}, Mechanisms I, III, and V ($M$ birth in association with bone rays, with alignment cues from the body and distal epithelial growth)
\begin{itemize}
\item Cell migration: We use the fin-specific values in Table~\ref{table:mig}.
\item Cell birth: We use the fin-specific values in Table~\ref{table:birth}. Additionally, we select locations for cell birth using the coordinates of the discretized bone rays in Eqn.~(\ref{bone}) (see Appendix~\ref{appendix:cellBirth} for details).
\item Cell death: We use the fin-specific values in Table~\ref{table:death}.
\item Skin growth: We do not scale cell positions with domain growth.
\item Switch parameters: In Eqn.~(\ref{eq:m}), we use $\zeta = -1$ cells. In Eqns.~(\ref{eq:xbirth}--\ref{eq:mbirth}), we use $d = 150$ $\mu$m.
\item Additional note on Fig.~\ref{Fig3}b: To calculate the distances that cells move per day, we consider the differences in their locations between consecutive days. In particular, the distance the $i$th $M$ cell moves in one day is $||\textbf{M}_i(t) - \textbf{M}_i (t + \Delta t_\text{mig,birth})|| + ||\textbf{M}_i(t+\Delta t_\text{mig,birth}) - \textbf{M}_i (t + 2\Delta t_\text{mig,birth})|| +||\textbf{M}_i(t +2\Delta t_\text{mig,birth}) - \textbf{M}_i (t + 1)||$, since $\Delta t_\text{mig,birth} = 1/3$ days in this simulation. If a new cell is born at position $\textbf{M}_j$ at, for example, time $t+\Delta t_\text{mig,birth}$, then we define the distance that cell agent moved between day $t$ and day $t+1$ as just  $||\textbf{M}_j(t+\Delta t_\text{mig,birth}) - \textbf{M}_j (t + 2\Delta t_\text{mig,birth})|| +||\textbf{M}_j(t +2\Delta t_\text{mig,birth}) - \textbf{M}_j (t + 1)||$.
\end{itemize}
\end{itemize}

}

\vspace{\baselineskip}
\noindent \textbf{Author Contributions:}{ A.V.\ and B.S.\ designed the study. All authors contributed to model development. M.R.A., D.C., N.C., B.D., and F.L. produced the growing fin domains; A.V., M.R.A., D.C., N.C., B.D., and F.L. performed simulations. A.V.\ wrote the paper; A.V.\ and B.S.\ contributed to revisions. All authors analyzed results and approved the manuscript.}

\vspace{\baselineskip}
\noindent \textbf{Acknowledgements:} The work of A.V.\ has been supported in part by the National Science Foundation (NSF) through DMS-1148284, DMS-1764421, and DMS-1440386; by the Mathematical Biosciences Institute; and by the Simons Foundation under grant no.\ 597491. M.R.A., D.C., N.C., B.D., and F.L.\ were supported by Brown University and ICERM through DMS-1439786. The work of B.S.\ was partially supported by the NSF through DMS-1408742 and DMS-1714429. We thank the N\"{u}sslein-Volhard lab for feedback during early model development, and are particularly grateful to April Dinwiddie for sharing her expertise on zebrafish fin patterns. We also recognize Emily Briggs, who contributed to earlier discussions on fin growth during an independent study with B.S.\ and A.V. We thank ICERM for hosting the undergraduate research component of this project.

\vspace{\baselineskip}
\noindent \textbf{Conflict of interest:} We declare we have no conflict of interest.

\bibliographystyle{spmpsci}  
\bibliography{bibliography}

\end{document}